\documentclass[a4paper,12pt]{article}
\usepackage{soul}
\usepackage[usenames,dvipsnames]{color}
\usepackage{jheppub}
\usepackage{esvect}
\usepackage{amsmath, amssymb, slashed, epsf, color, graphicx, latexsym}
\usepackage{tensor}
\usepackage{physics}
\usepackage{epsfig}
\usepackage{graphics}
\usepackage[]{natbib}
\usepackage[mathlines]{lineno}
\usepackage [english]{babel}
\usepackage [autostyle, english = american]{csquotes}
\MakeOuterQuote{"}

\begin{document}
\iffalse
\title{Study of near horizon dynamics of charged black hole in the presence of higher derivative correction}
\fi

\title{Equivalence of JT Gravity and Near-extremal Black Hole Dynamics in Higher Derivative Theory}

\author[a]{Nabamita Banerjee,}
\author[b]{Taniya Mandal,}
\author[a]{Arnab Rudra,}
\author[a]{and Muktajyoti Saha}
\affiliation[a]{Indian Institute of Science Education and Research Bhopal,
	Bhopal Bypass, Bhopal 462066, India.}
\affiliation[b]{School of Physics and Mandelstam Institute for Theoretical Physics, University of the Witwatersrand, Wits, 2050, South Africa}
%\affiliation[c]{Indian Institute of Science Education and Research Bhopal,
	%Bhopal Bypass, Bhopal 462066, India.}
	
\emailAdd{nabamita@iiserb.ac.in}
\emailAdd{taniya.mandal@wits.ac.za}
\emailAdd{rudra@iiserb.ac.in}
\emailAdd{muktajyoti17@iiserb.ac.in}

\abstract{Two derivative Jackiw–Teitelboim (JT) gravity theory captures the near-horizon dynamics of higher dimensional near-extremal black holes, which is governed by a Schwarzian action at the boundary in the near-horizon region. The partition function corresponding to this boundary action correctly gives the statistical entropy of the near-extremal black hole. In this paper, we study the thermodynamics of spherically symmetric four-dimensional near-extremal black holes in presence of arbitrary perturbative four derivative corrections. We find that the near-horizon dynamics is again captured by a JT-like action with a particular namely $R^2$ higher derivative modification. Effectively the theory is described by a boundary Schwarzian action which gets suitably modified due to the presence of the higher derivative interactions. Near-extremal entropy, free energy also get corrected accordingly.}
\maketitle
\section{Introduction}
Since decades, one of the major goals of theoretical physics is to understand the quantization of gravity. The simplest gravitational theory can be formulated in two dimensions. In two dimensions, the pure gravitational (in the absence of any matter field) action is topological i.e. the equations of motion are trivially satisfied and the action is proportional to the Euler Characteristic of the manifold. Thus to understand the two dimensional gravitational dynamics, we need to non-minimally couple matter fields with gravity\footnote{Minimally coupled matter field also does not bring any dynamics in the system.}. One such model is Callan-Giddings-Harvey-Strominger (CGHS) model in asymptotically flat space \cite{Callan:1992rs}, whereas another such model is Jackiw-Teitelboim (JT) \cite{Teitelboim:1983ux,Jackiw:1984je} theory, whose solution is asymptotically Anti de-Sitter(AdS) spacetime.

These gravitational theories, especially JT have drawn a great deal of attention in past few years. In the low energy limit, JT appears as gravitational dual to Sachdev-Ye-Kitaev (SYK) model, a solvable $0+1$ dimensional model of Majorana fermions with all possible random interactions  \cite{sy,k1,k2,Polchinski:2016xgd,Maldacena:2016hyu,Maldacena:2016upp} \footnote{Another interesting 2D non-minimally coupled gravity model is CGHS model that appears in the study of cosmology.}. For both SYK and JT, the time reparameterization symmetry is explicitly broken at the boundary of the spacetime, and the effective low energy dynamics is described by a Schwarzian action \cite{Jensen:2016pah,Sarosi:2017ykf}.
%{\color{Red}{Initially, this statement 'for both the theories' was meant fpr SYK and JT, but once this 'On the other hand, CGHS' line has been added, it seems, it changed the meaning of 'both the theories' as 'JT and CGHS'. does CGHS also given by Schwarzian?}} 
 In this article, we shall be interested in the JT gravity model. 

Pure $AdS_2$ is unique compared to any higher dimensional $AdS$ geometry. It lacks finite energy excitations in presence of gravitational backreaction as their existence modify the asymptotic boundary conditions \cite{Maldacena:1998uz}. For this reason, $AdS_2/CFT_1$ duality is not as well understood as higher dimensional AdS/CFT correspondence. On the contrary, JT gravity allows finite energy excitations in two dimensions, as the non trivial dilaton profile allows gravitational backreactions \cite{Almheiri:2014cka,Engelsoy:2016xyb}. Since the boundary conformal symmetry is slightly broken for JT, its solution is nearly $AdS_2$ or $NAdS_2$, also the boundary symmetry now is also nearly conformal or $NCFT_1$. Thus JT theory is a good candidate to study $NAdS_2/NCFT_1$ correspondence.

Recently it has been shown that JT gravity also describes the dynamics of the near horizon geometry of near-extremal black holes \cite{Maldacena:2016upp,Almheiri:2016fws}. The near horizon geometry of an extremal black hole has an $AdS_2$ part and possesses $SL(2,R)$ isometry which is weakly broken in the near-extremal limit, resulting in a $NAdS_2$ geometry. A JT-like action appears through the dimensional reduction of higher dimensional gravity. The connection between JT gravity and the near horizon dynamics of higher dimensional near-extremal charged and rotating black hole solution has been discussed in \cite{Nayak:2018qej,Moitra:2019bub,Moitra:2019xoj}. 
So far, the connection between the near-extremal black hole and JT gravity and other aspects of JT gravity like its connection with Matrix model has been proposed for two-derivative action \cite{Maldacena:2016upp,Almheiri:2016fws,Nayak:2018qej,Moitra:2019bub,Moitra:2019xoj,Kolekar:2018sba,Moitra:2021uiv,Saad:2019lba,Stanford:2019vob,Witten:2020wvy,Witten:2020ert}. In this paper, we check the validity of this proposal under small perturbations. We show that a JT-like action describes the dynamics of a near-extremal near horizon black hole physics appearing in four dimensional Einstein-Maxwell theory even with arbitrary small four derivative interactions involving metric and gauge field. The resultant JT theory gets modified by a precise two dimensional higher derivative interaction term.

Higher curvature terms occur at the low energy effective action in string theory as stringy corrections \cite{Zwiebach:1985uq,Gross:1986iv,Gross:1986mw,Myers:1987yn}. Properties of Einstein gravity have been widely studied in the presence of higher derivative terms e.g. in the context of hydrodynamics in AdS/CFT (on the CFT side, higher curvature terms correspond to finite t'Hooft coupling, finite $N$ corrections) \cite{Buchel:2008vz,Banerjee:2009wg}, entropy function formalism \cite{Sen:2007qy}, large dimensional black hole membrane paradigm \cite{Kar:2019kyz} etc. If higher derivative terms are not treated perturbatively, ghosts(unphysical states) may appear in the spectrum. This issue can be solved by considering particular combinations of the higher curvature terms, known as Lovelock terms, such that the equations of motion remain two derivative equations \cite{Lovelock:1971yv}. At four derivative level, terms of the Lovelock combination is known as the Gauss-Bonnet term: $R^2-4 R_{AB}R^{AB}+R_{ABCD}R^{ABCD}$. However in four-dimension, it is a topological term. Hence in this paper, we first consider an arbitrary quadratic curvature combination involving metric as $\lambda(\alpha_1 R^2+\alpha_2 R_{AB}R^{AB}+\alpha_3 R_{ABCD} R^{ABCD})$, where $\lambda$ is a dimensionless parameter controlling the strength of the curvature-squared terms, and $\alpha_1, \alpha_2$ and $\alpha_3$ are arbitrary coefficients of dimensions length square. To avoid ghosts (unphysical states), we treat these terms perturbatively. We consider the coupling constant $\lambda$ very small $\lambda \ll 1$ and take into account those terms which are linear in $\lambda$. Starting with a Einstein-Hilbert-Maxwell action in presence of such a quadratic curvature combination in four dimensions, we aim to get higher curvature modified JT-like action upon dimensional reduction, that will describe the dynamics of near-extremal black holes of the four dimensional theory. Our goal is legitimate, as the near-horizon geometry of an extremal black hole in the presence of higher derivative corrections still contains an $AdS_2$ factor, with a modified radius. Recently in \cite{Rathi:2021aaw}, a solution to quadratic corrected $2D$ gravity theory is presented and the corresponding boundary $CFT$ also has been studied. In the current manuscript we present the higher derivative corrected $2D$ gravity, whose dynamics is again captured by the boundary. Since we begin with the most generic four derivative curvature terms in four dimensions, one expects to get the most generic four derivative metric curvature corrections in two dimensional dilaton gravity theory, whose dynamics is still given by a Schwarzian action at the boundary. However the result of our computations shows that the corresponding JT theory only gets modified by $R^2$ term. The same feature continues even when we incorporate other possible four derivative interactions involving the metric and matter field.

Moving forward to a semi classical analysis, earlier it was thought that for a black hole with fixed charge, the difference of energies between extremal state and near-extremal state scales with temperature $T$ as $\Delta M=M-M_{ext}\sim \frac{T^2}{M_{gap}}$. As the required energy for a single Hawking quantum to radiate is of the order of the temperature $T$, thus when $T\lesssim M_{gap}$, the black hole would not have required energy to radiate. Hence the semi-classical analysis must breakdown (since a black hole is supposed to radiate with non-zero temperature). To make sense of semi classical analysis it was believed that there is a mass gap ~$M_{gap}$ between the extremal state and the closest near-extremal state in the mass spectrum for a fixed charged black hole, that prevents Hawking radiation for $T\lesssim M_{gap}$. Although the existence of a mass gap is well supported for supersymmetric black holes but in absence of supersymmetry, it is questionable. This issue has been resolved going beyond the semi classical regime, by considering the effects of quantum fluctuations. Recently in \cite{Iliesiu:2020qvm}, considering quantum fluctuations it has been showed that for a fixed charge Reissner-Nordstr\"{o}m black hole,\footnote{whose near-extremal dynamics is captured by the corresponding dimensional reduced JT action.} the $\Delta M $  acquires an extra term such that even at low temperature $\Delta M\sim \frac{3}{2}T$. As a consequence, the mass gap between the extremal and near-extremal black hole vanishes at low temperature, thus enabling Hawking radiation. This analysis provides a proper statistical description of black hole thermodynamics at low temperature. Hence, the entropy of near-extremal black hole possesses an extra correction, a logarithmic term in temperature which precisely omits the mass gap. Following their method, we also see the logarithmic correction in the entropy in presence of higher derivative curvature. 

Organisation of this paper is as follows: In section \ref{dilgrav}, we discuss the generic form of two-dimensional dilaton gravity theories with four-derivative corrections and also, we review some aspects of JT gravity which will be important for our computations. In section \ref{sec4} we present the modified JT theory that captures the low energy behavior of a spherically symmetric near-extremal black hole in four dimensional Einstein-Maxwell theory with four derivative metric interactions. This is the main result of this paper. We discuss the effects of more generic four derivative corrections involving the gauge field in section \ref{gen4derth}. We conclude the paper in section \ref{concl}. The paper contains six appendices, where we have included the relevant computational details.

\section{Dilaton gravity theories in two dimensions} \label{dilgrav}
\subsection{Some generalities }
The Einstein-Hilbert action in two dimensions is given by,
\begin{equation}
    S_{EH} = \frac{1}{16\pi G_2}\left[ \int d^2x \sqrt{-g} R + 2\int dx \sqrt{-h} K\right],
\end{equation}
where, $G_2$ is the dimensionless two dimensional Newton's constant. $R$ is the Ricci scalar constructed out of the two dimensional metric $g$. $h_{\mu\nu}=g_{\mu\nu} - n_\mu n_\nu$ is the induced metric on the one dimensional boundary having normal $n_\mu$ and extrinsic curvature $K=\nabla_\mu n^\mu$. This action is topological and proportional to the Euler Characteristic of the manifold. Variation of the action does not give any equation of motion since in two dimensions, Einstein tensor is identically zero for any metric.  
%\vspace{0.2cm}
%In $D$ dimensions, the number of independent components of the Riemann tensor is equal to $\frac{n(n+1)}{2}$ where $n=\frac{D(D-1)}{2}$. For $D=2$, the number of independent component is $1$, which is the Ricci scalar. It can be shown that,
%\begin{align}
 %   & R_{\alpha\beta\gamma\delta} = \frac{R}{2}(g_{\alpha\gamma}g_{\beta\delta}-g_{\alpha\delta}g_{\beta\gamma}) 
  %  & R_{\alpha\beta} = \frac{R}{2}g_{\alpha\beta} 
   % & G_{\alpha\beta} = R_{\alpha\beta} - \frac{1}{2}R g_{\alpha\beta} = 0
%\end{align}
%Because of this form of the Riemann tensor, variation of the Einstein-Hilbert action does not give any equation of motion for the metric. 
To make the theory non-trivial without deviating much, a non-minimally coupled scalar namely dilaton can be added. These dilaton-gravity theories in two dimensions have no local on-shell degrees of freedom. 
\iffalse
Dilaton-gravity theories arise from dimensional reduction of higher dimensional theories, where the dilaton is proportional to the volume of the compact space, over which the theory is reduced. In the context of this paper, we will be interested in JT theory and it's possible higher derivative corrections. JT gravity is a particular class of dilaton-gravity theory and it will be reviewed in section \ref{JTrev}. As discussed in appendix \ref{gendilact}, 
\fi
The most generic two derivative dilaton-gravity action \cite{Witten:2020ert} has the following form (\ref{gendilact}),
\begin{equation} \label{2ddlg}
    S = -\frac{1}{16\pi G_2}\left[ \int d^2x \sqrt{g}\left( \phi R + V(\phi)\right) + 2\int dx \sqrt{h} \phi K  \right].
\end{equation}
We get the action for JT gravity for a particular choice of $V(\phi)$ in the above action that will be described in the next section. The most generic action having terms up to four derivatives is given by \eqref{4derdilapp},

\begin{align}\label{4derdil}
    S = -\frac{1}{16\pi G_2}\int d^2x  \sqrt{g}\Bigg[\phi R + V(\phi) + \lambda W_1(\phi) (\nabla\phi)^4 & + \lambda W_2(\phi) R(\nabla\phi)^2 \nonumber \\
    & + \lambda W_3(\phi) R^2  \Bigg].
\end{align}

In higher dimensions, the four derivative corrected action contains terms quadratic in \textit{Riemann tensor}, \textit{Ricci tensor} and \textit{Ricci scalar}. But in two dimensions, the \textit{Riemann} tensor has only one component and that can be thought of as \textit{Ricci scalar} $R$, thus the higher derivative corrected action has terms involving $R$ and derivatives of the dilaton $\phi$\footnote{In principle, there could be more higher order terms \cite{Elizalde:1993ga}, but those can be absorbed by appropriate field redefinition (\ref{gendilact}).}.

\subsection{A Brief overview of the Jackiw-Teitelboim (JT) gravity} \label{JTrev}
In this section, following \cite{Teitelboim:1983ux,Jackiw:1984je, Maldacena:2016upp}, we shall briefly review some important aspects of Jackiw-Teitelboim gravity model that would be required for our present study. Experts may skip this section and directly study from section \ref{sec4}.

The Jackiw-Teitelboim or JT gravity model is a specific kind of two dimensional dilaton gravity theory, described by the action \eqref{2ddlg} with a linear dilaton potential $V(\phi) = -\Lambda_2\phi$. The action in Lorentzian signature is given by,
\begin{align}
    & S_{JT} = \frac{1}{16\pi G_2}\left[ \int d^2x \sqrt{-g} \phi \left( R -\Lambda_2\right) + 2\int dx \sqrt{-h} \phi K  \right] \label{JT},
\end{align}
where $\Lambda_2$ is the two dimensional cosmological constant. In the full action, we add the topological Einstein-Hilbert term with the JT part, thus the full action is $S = \phi_0 S_{EH} + S_{JT}$. Here $\phi_0$ is a constant, whose implications will be clarified in the next subsection. The equation of motion of the scalar $\phi$ sets the curvature to a constant value $\Lambda_2$, thus depending on the sign of $\Lambda_2$ we get asymptotically $dS$ or $AdS$ spacetime. In this work we consider $\Lambda_2 = -\frac{2}{L_2^2}$, for which the spacetime is asymptotically $AdS_2$, having a length scale $L_2$. Equations of motion of the scalar and the metric are respectively,
\begin{eqnarray}
    && R + \frac{2}{L_2^2} = 0 \label{dileom},\\
    && \nabla_\mu\nabla_\nu\phi - \nabla^2\phi g_{\mu\nu} + \frac{\phi}{L_2^2} g_{\mu\nu} = 0 \label{meteom}.
\end{eqnarray}
The solutions of the equation \eqref{dileom} are locally $AdS_2$. The generic form of the solution is \cite{Cadoni:1999ja},
\begin{align}
    ds^2 = - \left(\frac{r^2}{L_2^2}-m^2\right)dt_L^2 + \frac{dr^2}{\frac{r^2}{L_2^2}-m^2} \label{aads2}.
\end{align}
For $m^2 \geq 0$, the spacetime has horizons. These solutions, characterized by $m^2$, cover patches of $AdS_2$ but they can be maximally extended to global $AdS_2$ using coordinate transformations\footnote{the same also holds true for the euclidean solution.}. Therefore, they describe the same spacetime and these solutions are physically equivalent. 
%where the $AdS_2$ length $L_2$ is related to the two dimensional cosmological constant as: $\Lambda_2 = -\frac{2}{L_2^2}$. 

The dilaton can take non-negative values only\footnote{This statement will be justified later, where the v.e.v. of the dilaton will be related to the radius of the compact dimensions.}. From \eqref{meteom}, the only constant solution of the dilaton is $\phi=0$. When $\phi$ is constant, the extension from \eqref{aads2} to global $AdS_2$ is well defined. But for a varying solution e.g. $\phi\sim r$, the metric \eqref{aads2} cannot be maximally extended since the dilaton becomes negative in the region $r<0$. The requirement of its positivity prevents the maximal extension of the spacetime and hence it should be restricted to $\phi \geq 0$ region. With the non-trivial dilaton, the full solutions of the theory (i.e. metric and dilaton) are nonequivalent depending on $m^2$. We will be interested in "black hole" solutions ($m^2\geq 0$) i.e. solutions having horizons and temperature.  For these solutions, the metric can be put into Poincar\'{e} $AdS_2$ form in Euclidean signature\footnote{Since we are interested in the Euclidean path integral, we will consider Euclidean time $t_L\rightarrow it$.}, 
 \begin{eqnarray}
    ds^2 = g_{\mu\nu}dx^\mu dx^\nu = \frac{L_2^2}{z^2}(dt^2 + dz^2) \label{adsmet}.
\end{eqnarray}
The generic solution to the dilaton equation \eqref{meteom} is given by \cite{Maldacena:2016upp},
\begin{eqnarray}\label{dilsol}
    \phi (t,z) = \frac{K_1 + K_2 t + K_3 (t^2 + z^2)}{z} ,
\end{eqnarray} where $K_1, K_2$ and $K_3$ are integration constants and $\phi \geq 0$. This non-trivial dilaton solution breaks the $SL(2,R)$ isometry of $AdS_2$ to $U(1)$, generated by the vector field $X^\mu = \epsilon^{\mu\nu}\nabla_\nu\phi$.

\subsubsection{Breaking of Boundary Time reparametrization symmetry}
The spacetime \eqref{adsmet} has a one-dimensional timelike boundary at $z=0$ \cite{Maldacena:2016upp}. Since the fields diverge near the boundary, we cut the spacetime along a curve $\mathcal{C}: (t(u),z(u))$, parametrized by boundary time coordinate $u\in [0,\beta_0]$. We fix the length of the boundary curve by imposing Dirichlet boundary condition on the metric,
\begin{eqnarray}
    h = g|_\mathcal{C} = \frac{L_2^2}{\varepsilon^2}, \label{metbc}
\end{eqnarray}
where the length of the boundary curve is given by,
\begin{align}
    l_{bdy} =  \int_{\mathcal{C}} du \sqrt{h} = \frac{\beta_0 L_2}{\varepsilon}.
\end{align}
Since the metric \eqref{adsmet} should satisfy the boundary condition \eqref{metbc}, we find that  
\begin{eqnarray}
     g_{uu} &=& \frac{L_2^2}{z^2} (t'^2 + z'^2) = \frac{L_2^2}{\varepsilon^2}.
\end{eqnarray}
For small $\varepsilon$, solution to this equation is given by, 
\begin{equation}
      z(u) \simeq \varepsilon t'(u), \label{zb}
\end{equation}
where, the primes denote derivatives w.r.t $u$. Clearly, the shape of the boundary curve is determined by the function $t(u)$ such that $t'(u)>0$. 

Although locally the spacetime is $AdS_2$, different choices of the function $t(u)$ correspond to different geometries since the cut out shape depends on the form of the function. For a constant dilaton solution i.e. $\phi=0$, each of these geometries correspond to the ground state since the Einstein-Hilbert action is topological and it takes the same value for any cut out shape given by $t(u)$, for any value of the constant $\phi_0$. Thus we have a time reparametrization symmetry near the boundary, whose generators are as follows,
\begin{eqnarray}
  \xi^t = \chi(t), \qquad \xi^z = z\frac{d\chi(t)}{dt}.  
\end{eqnarray}
This is the realization of asymptotic symmetry of $AdS_2$. These reparametrizations map one boundary curve to another e.g. $t \rightarrow t + \chi (t)$, though not all of them are different. Since the isometry group of $AdS_2$ is $SL(2,R)$, functions $t(u)$ and $\hat{t}(u)$, which are related by $SL(2,R)$ transformations \footnote{The transformation relation is : \begin{align}
    \hat{t}(u) = \frac{\alpha t(u) + \beta}{\gamma t(u) + \delta}; \qquad \alpha\delta - \gamma\beta = 1.
\end{align}}, they correspond to the same cutout shape.

The ground state geometries are then characterized by different functions $t(u)$ up to $SL(2,R)$ identification. The time reparametrization symmetry is spontaneously broken to $SL(2,R)$ by the choice of $AdS_2$ vacuum. 

On the contrary the time reparametrization symmetry is explicitly broken in the JT action. Boundary condition for the dilaton is taken as
\begin{align}
    \phi|_{\mathcal{C}} = \frac{\phi_b}{\varepsilon}, \label{dilbc}
\end{align}
where $\phi_b$ is a dimensionful number specifying the boundary value of the dilaton. This $\phi_b$ characterizes the scale of symmetry breaking. We consider $\phi_b \ll \phi_0 L_2$ such that the symmetry is slightly broken. The generic solution \eqref{dilsol} of the dilaton should satisfy this boundary condition. Thus using \eqref{zb} and \eqref{dilbc},  we find: 

\iffalse
\begin{align}
    \phi = \frac{K_1 + K_2 t + K_3 (t^2 + \varepsilon^2 t'^2)}{\varepsilon t'}.
\end{align}
Imposing the boundary condition \eqref{dilbc}, we find
\fi

\begin{equation}
    \frac{K_1 + K_2 t + K_3 t^2}{t'} = \phi_b. \label{mode_sol}
\end{equation}
Solving this equation for the function $t(u)$, we get the classical shape of the boundary.

\subsubsection{Effective boundary action}\label{EPI}
The Euclidean path integral corresponding to the action $S=\phi_0 S_{EH}+S_{JT}$ is as follows,
\begin{align}
    Z = \int \mathcal{D}\phi\mathcal{D}g_{\mu\nu}\text{e}^{-\phi_0 S_{EH}-S_{JT}}.
\end{align}
Using Gauss-Bonnet theorem, the Einstein-Hilbert action is given by\footnote{This term contributes to the entropy of the higher dimensional extremal black hole, whereas the action with dynamical dilaton contributes to the entropy beyond extremality.},
\begin{align}
    S_{EH} = -\frac{1}{8G}\chi(M),
\end{align}
where $\chi(M)=2-2g-n$ is the Euler Characteristic of the manifold $M$ of genus $g$ with $n$ boundaries. For large value of $\phi_0$, the dominant contribution to the partition function comes from disk topology with one boundary i.e. $g=0, n=1$ and higher genus manifolds with multiple boundaries have exponentially suppressed contribution. Thus the partition function takes the following form
\begin{align}
    Z = \text{e}^{\frac{\phi_0}{8G}(2-2g-n)}\int \mathcal{D}\phi\mathcal{D}g_{\mu\nu}\text{e}^{-S_{JT}} = \text{e}^{\frac{\phi_0}{8G}(2-2g-n)}Z_{JT}. \label{PI}
\end{align}

Since there are no local bulk excitations, the dynamics is effectively governed by a boundary action. This can be understood by integrating out the dilaton field in the JT path integral\footnote{We will discuss only the disk topology here.} corresponding to action \eqref{JT} for $\Lambda_2=-\frac{2}{L_2^2}$,
\begin{align}
    Z_{JT} = \int \mathcal{D}\phi\mathcal{D}g_{\mu\nu}\text{exp}\left[\frac{1}{16\pi G_2} \int d^2x \sqrt{g} \phi \left( R +\frac{2}{L_2^2}\right) + \frac{1}{8\pi G_2}\int_{\phi=\phi_b} dx \sqrt{h} \phi K  \right]. 
\end{align}
The dilaton acts as a Lagrange multiplier in this integral. Integrating the dilaton along the imaginary line with Dirichlet boundary condition at the boundary, we get a delta function, hence
\begin{align}
    Z_{JT} = \int\mathcal{D}g_{\mu\nu}\delta\left( R +\frac{2}{L_2^2}\right)\text{exp}\left[\frac{1}{8\pi G_2}\int_{\phi=\phi_b} dx \sqrt{h} \phi K  \right] \label{PI1}.
\end{align}
Using the delta function, the above path integral reduces to sum over field configurations for which the condition $R +\frac{2}{L_2^2}=0$ is met.

The effective theory can also be obtained by simply imposing the $\phi$ equation of motion \eqref{dileom} in the action. The bulk term of \eqref{JT} vanishes and it reduces to the boundary term involving the extrinsic curvature at the boundary $\mathcal{C}: (t(u),z(u))$,
\begin{align}
    I = -\frac{1}{8\pi G_2} \int_\mathcal{C} du \frac{L_2}{\varepsilon} \frac{\phi_b}{\varepsilon} K,
\end{align}
where the extrinsic curvature $K$ is given by
\begin{align}
    K = \frac{t'(t'^2+z'^2+zz'')-zz't''}{L_2(t'^2+z'^2)^{3/2}}.
\end{align}
Using the boundary condition \eqref{zb} we get,
\begin{align}
    K = \frac{1}{L_2}\left(1 + \varepsilon^2 \text{Sch}[t(u),u]\right).
\end{align}
Here $\text{Sch}[t(u),u]$ is the Schwarzian derivative of the time reparametrization mode $t(u)$, defined as
\begin{align}
    \text{Sch}[t(u),u] = -\frac{1}{2}\frac{t''^2}{t'^2} + \left(\frac{t''}{t'}\right)'. \label{Sch}
\end{align}
Then the effective action takes the following form\footnote{The constant term in the expression of $K$ diverges as $\varepsilon\rightarrow 0$, but it can be removed by adding a counter term $\frac{1}{8\pi G_2 L_2}\int dx \sqrt{h}\phi$ to the action. The resulting effective action is independent of $\varepsilon$.}
\begin{align}
    I[t(u)] =  -\frac{\phi_b}{8\pi G_2} \int du\text{Sch}[t(u),u]. \label{Sch_action}
\end{align}
Evidently, the time reparametrization modes $t(u)$ acquire an action. Variation of the action gives the classical value of $t(u)$. It should be noted that the action is $SL(2,R)$ invariant. This is a gauge freedom in the choice of modes since different $t(u)$ related by $SL(2,R)$ transformations correspond to the same configuration. 
Hence, the path integral \eqref{PI1} gets contribution from the boundary modes $t(u)$ up to $SL(2,R)$ identification with an appropriate choice of boundary diffeomorphism invariant measure \cite{Saad:2019lba} and is given by
\begin{align}
    Z_{JT}(\beta_0) = \int \frac{d\mu(t)}{SL(2,R)}\text{exp}\left(\Bar{\phi}_b \int^{\beta_0}_0 du\text{Sch}[t(u),u]\right). \label{SchPI}
\end{align}
Here $\beta_0$ is the range of the time direction and corresponds to the inverse temperature $\beta_0 = 1/T_0$. $\bar{\phi}_b$ is the coupling constant of the Schwarzian theory, given as follows
\begin{align}
   \bar{\phi}_b =\frac{\phi_b}{8\pi G_2}. \label{sch_coup}
\end{align}
The functions $t(u)$ are elements of $\text{Diff}(S^1)/SL(2,R)$, which is a symplectic manifold. Since the path integral is over this manifold, the measure $d\mu(t)$ can be chosen to be the natural measure induced from the symplectic form given by \cite{Stanford:2017thb}
\begin{align}
    d\mu(t) = \prod \frac{dt}{t'}. \label{measure}
\end{align}
Furthermore in \eqref{SchPI} $\frac{d\mu(t)}{SL(2,R)}$ implies measure $d\mu(t)$ modded by $SL(2,R)$.  Since the boundary action is $SL(2,R)$ invariant, the modes related by these gauge transformations should be identified in the measure.

\subsubsection{Evaluation of exact partition function}
\label{NE_stat}
\iffalse
Under a field redefinition $t = \tan{\frac{\tau}{2}}$ and considering a constant boundary value of the dilaton, the action takes the form:
\begin{align}
    I[\tau(u)] = -\frac{\phi_b}{8\pi G_2}\int du \left[\text{Sch}[\tau(u),u] + \frac{1}{2}\tau'(u)^2\right]. \label{Schact}
\end{align}
A solution to the equation of motion derived from this action is $\tau(u) = \frac{2\pi}{\beta} u$. 
\fi

The classical solution of the boundary action \eqref{Sch_action} is of the form:
\begin{align}
    t(u) = \tan{\left(\frac{\pi u}{\beta_0}\right)}. 
\end{align}
Doing a saddle point analysis, the dominant contribution to the partition function above extremality is obtained from the on-shell action,
\begin{align}
    -& \ln{\delta\mathcal{Z}(T_0)} \simeq I_{\text{on-shell}} = -2\pi^2\bar{\phi}_b T_0,
\end{align}
whereas the free energy, mass, and entropy above extremality are,
\begin{eqnarray}
     \delta F &=& -T_0 \ln{\delta\mathcal{Z}(T_0)} = -2\pi^2\bar{\phi}_b T_0^2,  \\
     \delta S &=& -\frac{\partial \delta F}{\partial T_0}  = 4\pi^2\bar{\phi}_b T_0, \\
     \delta M &=& \delta F + T_0\delta S = 2\pi^2\bar{\phi}_b T_0^2.
\end{eqnarray}
The ground state is given by the $T_0\rightarrow 0$ limit such that $\delta F=0$. In this limit, the time reparametrization symmetry gets restored and this  configuration is that of an extremal black hole.

It was shown in \cite{Stanford:2017thb} that the partition function is one-loop exact. Hence it is enough to consider the contribution of small fluctuations around the classical solution,
\begin{align}
    t(u) = \tan{\frac{\pi}{\beta_0}(u + \rho(u))}.
\end{align}
Expanding the action \eqref{Sch_action} to quadratic order in fluctuations, we find
\begin{align}
    I[\rho(u)] =& -\bar{\phi}_b\int^{\beta_0}_0 du \left[ \frac{2\pi^2}{\beta_0^2} + \left(\frac{4\pi^2}{\beta_0^2}\rho' + \rho'''\right) + \left(\frac{2\pi^2}{\beta_0^2}\rho'^2 - \frac{3}{2}\rho''^2 - \rho'\rho''' \right)\right], \nonumber \\
    =& -\frac{2\pi^2\bar{\phi}_b}{\beta_0} - \bar{\phi}_b\int^{\beta_0}_0 du\left(\frac{2\pi^2}{\beta_0^2}\rho'^2 - \frac{3}{2}\rho''^2 - \rho'\rho''' \right).
\end{align}
The linear order term drops out due to periodicity of the boundary time coordinate. The fluctuation parameter $\rho(u)$ can be decomposed in Fourier modes,
\begin{align}
    \rho(u) = \sum_{\abs{n}\geq 2} c_n \text{e}^{-i\frac{2\pi n}{\beta_0}u}.
\end{align}
Since $\rho(u)$ is real, we have $c^*_n = c_{-n}$. The measure \eqref{measure} now depends on these modes, given by
\begin{align}
    d\mu = \frac{(2\pi)^3}{\beta_0^2}\prod_{n\geq 2} (n^3-n)dc_n dc^*_n.
\end{align}
The modes corresponding to $n=0,\pm1$ are removed from the measure by construction since these correspond to fluctuations due to infinitesimal $SL(2,R)$ transformations. For this quadratic action, the partition function takes the form of a Gaussian integral in fluctuations,
\begin{align}
    Z_{JT}(\beta_0) = \text{e}^{\frac{2\pi^2\phi_b}{\beta_0}} \frac{(2\pi)^3}{\beta_0^2}\prod_{n\geq 2} (n^3-n) \int dc_n  dc^*_n \text{exp}\left(\frac{16\pi^4\bar{\phi}_b}{\beta_0^3}(n^2-n^4)c_n c^*_n\right).
\end{align}
The modes $c_n$ can be integrated out to give,
\begin{align}
    Z_{JT}(T_0) \simeq \left(\bar{\phi}_b T_0\right)^{3/2}\text{e}^{2\pi^2\bar{\phi}_b T_0},
\end{align}
here we have substituted $\beta_0 = 1/T_0$. The corresponding thermodynamic quantities are expressed in terms of the coupling $\bar{\phi}_b$ and the temperature parameter $T_0$ as follows\footnote{From higher dimensional perspective, these quantities with appropriate choice of $\phi_b$ and $G_2$ capture the thermodynamics of a near-extremal black hole with temperature $T_0$.},
\begin{align}
    & \delta F  = -2\pi^2\bar{\phi}_b T_0^2 - \frac{3T_0}{2}\log{\left(\bar{\phi}_bT_0\right)},  \\
    & \delta S = \frac{3}{2} + 4\pi^2\bar{\phi}_bT_0 + \frac{3}{2}\log{\left(\bar{\phi}_bT_0\right)}, \\
    & \delta M  = 2\pi^2\bar{\phi}_b T_0^2 + \frac{3T_0}{2}.
\end{align}

\section{Spherically symmetric near-extremal black hole in higher derivative theory} \label{sec4}
The near-horizon geometry of a spherically symmetric extremal black hole in $d+2$ dimensions factorises in $AdS_2\times S^d$. It has been shown in \cite{Cadoni:1993rn,Cadoni:1994uf} that if we dimensionally reduce a higher dimensional Einstein-Hilbert-Maxwell theory on a spherically symmetric background and restrict ourselves to the s-wave sector, the effective lower dimensional theory of the massless fields consist of a metric, gauge field and a dilaton. The classical solution of the effective theory rightly reproduces the extremal near horizon geometry of the higher dimensional theory. Fluctuating this effective theory around the classical solution results in JT gravity and it corresponds to the dynamics of near-extremal black holes in the higher dimensions. The partition function of the JT gravity correctly gives the entropy difference beyond that of an extremal black hole. In this paper we
show that the above equivalence of the near-extremal black hole dynamics to JT gravity holds even in presence of a small perturbative correction to the higher dimensional theory. In particular we
study the JT equivalent model that appears after the dimensional reduction of four dimensional Einstein-Hilbert-Maxwell theory in the presence of four derivative metric interactions on a spherically symmetric background. We begin with an action in four dimensional spacetime with a negative cosmological constant in presence of arbitrary four derivative corrections. The bulk action is given by,  
\begin{eqnarray}\label{4Dbulk1}
     \hat{S}_{bulk}&& = \frac{1}{16\pi G} \int d^4x \sqrt{-\hat{g}} \Big( \hat{R} - 2\Lambda -\hat{F}_{AB}\hat{F}^{AB} + \lambda\alpha_1\hat{R}^2 + \lambda\alpha_2\hat{R}_{AB}\hat{R}^{AB} \nonumber\\&&+ \lambda\alpha_3\hat{R}_{ABCD}\hat{R}^{ABCD}  \Big) .
\end{eqnarray}
Here $G$ is the four dimensional Newton's constant with dimension of length square and the higher derivative coefficients $\alpha_1, \alpha_2, \alpha_3$ also have the dimension of length square. $\lambda$ is a dimensionless number controlling the strengths of the higher derivative terms.
The Gauss-Bonnet combination $\hat{R}_{GB}^2=\hat{R}^2-4\hat{R}_{AB}\hat{R}^{AB}+\hat{R}_{ABCD}\hat{R}^{ABCD}$ is a topological term in four dimensions and does not contribute to the equations of motion. Motivated by this, the higher derivative part of the action can be rewritten as,
\begin{eqnarray}
    \alpha_1\hat{R}^2 + \alpha_2\hat{R}_{AB}\hat{R}^{AB} + \alpha_3\hat{R}_{ABCD}\hat{R}^{ABCD} = \tilde{\alpha}_1\hat{R}^2 + \tilde{\alpha}_2\hat{R}_{AB}\hat{R}^{AB} + \alpha_3\hat{R}_{GB}^2,
\end{eqnarray}
where we have used $\tilde{\alpha}_1 = \alpha_1-\alpha_3$ and $\tilde{\alpha}_2 = \alpha_2+4\alpha_3$. The rewritten bulk action is,
\begin{eqnarray}\label{4Dbulk}
    \hat{S}_{bulk} && =\frac{1}{16\pi G} \int d^4x \sqrt{-\hat{g}} \Big( \hat{R} - 2\Lambda -\hat{F}_{AB}\hat{F}^{AB} + \lambda(\tilde{\alpha}_1\hat{R}^2 + \tilde{\alpha}_2\hat{R}_{AB}\hat{R}^{AB} \nonumber\\
      && + \alpha_3\hat{R}_{GB}^2) \Big).
\end{eqnarray}
The boundary action, consistent with Dirichlet boundary condition on the metric, is given by \cite{Cremonini:2009ih},
\begin{eqnarray}\label{4Dbdy}
     \hat{S}_{bdy} &=& \frac{1}{8\pi G} \int d^3x \sqrt{-\hat{h}} \Bigg[ \left( 1-\lambda\frac{24}{L^2}\tilde{\alpha}_1 -\lambda\frac{6}{L^2}\tilde{\alpha}_2 \right)\hat{K} + \lambda\tilde{\alpha}_2 (-\hat{K}\hat{F}^2+2\hat{K}\hat{F}^{AB}\tensor{\hat{F}}{^C_B}\hat{n}_A \hat{n}_C  \nonumber \\
    && + 2\hat{K}_{ab}\hat{F}^{aD}\tensor{\hat{F}}{^b_D}) + 2\lambda\alpha_3 (\hat{J} - 2\hat{G}^{(3)}_{ab}\hat{K}^{ab}) \Bigg],
\end{eqnarray}
where $\hat{J}$ is the trace of: 
\begin{equation}
    \hat{J}_{ab} = \frac{1}{3} (2\hat{K}\hat{K}_{ac}\hat{K}^c_b + \hat{K}_{cd}\hat{K}^{cd}\hat{K}_{ab} -2\hat{K}_{ac}\hat{K}^{cd}\hat{K}_{db} -\hat{K}^2\hat{K}_{ab}) \label{J_ab}.
\end{equation}
The four dimensional cosmological constant $\Lambda$ is given in terms of the four dimensional $AdS$ radius $L$ as follows
\begin{equation}
    \Lambda = -\frac{3}{L^2}. 
\end{equation}

The boundary term has explicit dependence on the gauge field strength. Also we are interested in fixed electrically charged solutions, hence we cannot apply Dirichlet boundary condition on the gauge field. Instead we impose the boundary condition $\delta(n_A F^{A b}) = 0$. For a well-defined variational principle for the gauge field, consistent with this boundary condition, we need to add the following Maxwell boundary term \cite{Braden:1990hw,Hawking:1995ap,Chamblin:1999tk},
\begin{align}
    \hat{S}_{Maxb}=\frac{1}{4\pi G}\int d^3x\sqrt{-\hat{h}}\hat{n}_C \hat{F}^{CD}\hat{A}_D. \label{Max_bdy}
\end{align}
Adding \eqref{4Dbulk}, \eqref{4Dbdy}, and \eqref{Max_bdy}, the full four dimensional action is given by
\begin{align}
    \hat{S} = \hat{S}_{bulk} + \hat{S}_{bdy} + \hat{S}_{Maxb}. \label{4D_act}
\end{align}

\iffalse
The boundary is located at asymptotic infinity and it is a timelike boundary. The induced metric:
\begin{align}
    \hat{h}_{AB} = \hat{g}_{AB} - \hat{n}_A\hat{n}_B
\end{align}
\fi

\subsection{Attractor solutions}
In four dimensions, the near horizon geometry of a spherically symmetric extremal black hole takes the form $AdS_2\times S^2$. %The $AdS_2$ and $S^2$ radii can be obtained using Sen's entropy function formalism \cite{Sen:2007qy}.
At the level of two-derivatives, the near horizon values of the $S^2$ radius and that of $AdS_2$ of an extremal black hole are given by \cite{Sen:2007qy},
\begin{align}
    & \Phi_0^2 = r_h^2 = \frac{L^2}{6}\left(\sqrt{1+\frac{12Q^2}{L^2}}-1\right), \label{leadphi0} \\
   & L_2^2 = \frac{L^2}{6}\left(1-\frac{1}{\sqrt{1+\frac{12Q^2}{L^2}}}\right) = \frac{\Phi_0^2}{1+\frac{6\Phi_0^2}{L^2}},
\end{align}
where we have considered an electrically charged solution with charge $Q$. These values get corrected in the presence of higher derivative corrections though the form of the near horizon geometry of the extremal black hole remains the same as before. The   
corrected attractor values of $S^2$ and $AdS_2$ radii are as follows\footnote{The attractor values are derived in \ref{entfn} using entropy function formalism.},
\begin{align}
    & \tilde{\Phi}_0 = \tilde{r}_h = \Phi_0 + \lambda(2\tilde{\alpha}_1 + \tilde{\alpha}_2)\frac{L^2+3\Phi_0^2}{L^2+6\Phi_0^2}\frac{6\Phi_0}{L^2}, \label{extS2} \\
   & \tilde{L}_2 = L_2 + \lambda(2\tilde{\alpha}_1 + \tilde{\alpha}_2)\frac{L^2+3\Phi_0^2}{(L^2+6\Phi_0^2)^{5/2}}6\Phi_0L. \label{extAdS2}
\end{align}
Here we have rewritten the charge parameter $Q$ in terms of the unmodified $S^2$ radius using \eqref{leadphi0}. It can be noted that if we set the particular combination of coefficients $(2\tilde{\alpha}_1 + \tilde{\alpha}_2)$ to zero, the extremal solution remains unchanged. It is a consequence of the near-horizon configuration of the extremal solution, which is a product of maximally symmetric spaces. Although the extremal solution does not change under the condition $(2\tilde{\alpha}_1 + \tilde{\alpha}_2) = 0$, non-extremal solutions get modified.

\subsection{Dimensional reduction of four dimensional theory over \texorpdfstring{$S^2$}{text}} 

In this section, we reduce the four dimensional theory \eqref{4D_act} to two dimensions in the s-wave sector of massless fields\footnote{In principle, for a consistent Kaluza Klein reduction, all the massless modes should be included which involve $SO(3)$ gauge fields \cite{pope, Michelson:1999kn}. In appendix \ref{KK}, we argue why these modes are not required in the present considerations.}. We consider spherically symmetric configurations in four dimensions having the following form
\begin{align}
    & ds^2 = \tilde{g}_{\alpha\beta}dx^\alpha dx^\beta + \Phi^2(x^\alpha) (d\theta^2 + \sin^2\theta d\varphi^2) \label{S^2_red}, \quad x^\alpha=t,r, \\
    &\hat{A}_{\mu} = \tilde{A}_\mu, \quad \hat{A}_i = 0.
\end{align}
The form of the gauge field corresponds to electrically charged configurations. However, magnetic charge can also be considered since these are related by duality transformations. Using the above ansatz, we integrate the four dimensional action \eqref{4D_act} over the angular coordinates $\theta$ and $\varphi$. The reduced action involves the scalar $\Phi$, the two-dimensional metric $\tilde{g}_{\mu\nu}$, the two-dimensional abelian gauge field $\tilde{A}_\mu$ and their derivatives. In two dimensions, the curvature tensors have only one independent component which is the Ricci scalar $\tilde{R}$. Also, the extrinsic curvature tensor can be written in terms of its trace. In particular, in two dimensions we have,
\begin{align}
    & \tilde{R}_{\alpha\beta\gamma\delta} = \frac{\tilde{R}}{2}(\tilde{g}_{\alpha\gamma}\tilde{g}_{\beta\delta}-\tilde{g}_{\alpha\delta}\tilde{g}_{\beta\gamma}), \\
    & \tilde{R}_{\alpha\beta} = \frac{\tilde{R}}{2}\tilde{g}_{\alpha\beta}, \\
    & \tilde{K}_{\alpha\beta} = \tilde{K}\tilde{h}_{\alpha\beta},
\end{align} 
where, $\tilde{K}$ is the trace of extrinsic curvature and $\tilde{h}_{\alpha\beta}=\tilde{g}_{\alpha\beta} - \tilde{n}_{\alpha}\tilde{n}_{\beta}$ is the induced metric on the one-dimensional timelike boundary having a normal $\tilde{n}_\mu$. Using these relations we find the full action %(i.e. containing both bulk and boundary parts)
after dimensional reduction,
\begin{align}
    \tilde{S} = \Tilde{S}^{(0)} + \Tilde{S}^{\tilde{\alpha}} + \Tilde{S}^{GB}. \label{2D_act}
\end{align}
Here we have split the reduced action into three parts for convenience. $\Tilde{S}^{(0)}$ is the $\lambda^0$ order action i.e. it contains the two-derivative terms only. The $\lambda^1$ order action is further split into two parts: $\Tilde{S}^{\tilde{\alpha}}$ (depending on the couplings $\tilde{\alpha}_1$ and $\tilde{\alpha}_2$) and $\Tilde{S}^{GB}$ (depending on the coupling $\alpha_3$ of the Gauss-Bonnet term). The parts are as follows:

\begin{itemize}

\item Reduced two derivative part:
\begin{align}
    \Tilde{S}^{(0)} = & \frac{1}{4G}\int d^2x \sqrt{-\tilde{g}} \left[\Phi^2 (\tilde{R} -2\Lambda - \tilde{F}^2) + 2 + 2(\tilde{\nabla}\Phi)^2 \right] \nonumber \\
    & + \frac{1}{2G}\int dx \sqrt{-\tilde{h}} \Phi^2 (\tilde{K} + 2\tilde{n}_\alpha \tilde{F}^{\alpha\beta}\tilde{A}_\beta ),
\end{align}
\item Reduced higher derivative parts with arbitrary coefficients:
\begin{eqnarray}
    && \Tilde{S}^{\tilde{\alpha}}  = \frac{\lambda}{4G} \int d^2x \sqrt{-\tilde{g}}\Bigg[\left(4\tilde{\alpha}_1+2\tilde{\alpha}_2\right)\frac{1}{\Phi^2} + 4\tilde{\alpha}_1 \tilde{R} - \left(16\tilde{\alpha}_1+4\tilde{\alpha}_2\right)\frac{\tilde{\nabla}^2\Phi}{\Phi}- 4\tilde{\alpha}_1\tilde{R} (\tilde{\nabla}\Phi)^2\nonumber \\
    &&-\left(8\tilde{\alpha}_1+4\tilde{\alpha}_2\right)\frac{(\tilde{\nabla}\Phi)^2}{\Phi^2}   + \left(\tilde{\alpha}_1+\frac{\tilde{\alpha}_2}{2}\right)\tilde{R}^2\Phi^2 -\left(8\tilde{\alpha}_1+2\tilde{\alpha}_2\right)\tilde{R}\Phi\tilde{\nabla}^2\Phi  + 4\tilde{\alpha}_2(\tilde{\nabla}^\alpha\tilde{\nabla}^\beta \Phi)^2  \nonumber \\
    &&  +\left(16\tilde{\alpha}_1+2\tilde{\alpha}_2\right)(\tilde{\nabla}^2\Phi)^2+\left(4\tilde{\alpha}_1+2\tilde{\alpha}_2\right)\frac{(\tilde{\nabla} \Phi)^4}{\Phi^2} + \left(16\tilde{\alpha}_1+4\tilde{\alpha}_2\right)\frac{\tilde{\nabla}^2\Phi (\tilde{\nabla}\Phi)^2}{\Phi}  \Bigg] \nonumber \\
    && + \frac{\lambda}{2G} \int dx \sqrt{-\tilde{h}} \Bigg[\left(-\frac{24\tilde{\alpha}_1}{L^2}-\frac{6\tilde{\alpha}_2}{L^2}\right)\left(\Phi^2\tilde{K}+2\Phi\tilde{n}\cdot\tilde{\nabla} \Phi\right) + \tilde{\alpha}_2 \Big(\Phi^2 \tilde{K} \tilde{F}^2 \nonumber \\
    &&+2\Phi\tilde{n}\cdot\tilde{\nabla} \Phi \left(2\tilde{F}^{\alpha\gamma}\tensor{\tilde{F}}{^\beta_\gamma}\tilde{n}_\alpha \tilde{n}_\beta - \tilde{F}^2\right) \Big)\Bigg],
\end{eqnarray}
\item Reduced Gauss-Bonnet part: 
\begin{eqnarray}
    && \Tilde{S}^{GB}  = \frac{\lambda\alpha_3}{4G} \int d^2x \sqrt{-\tilde{g}}\Bigg[ 4\tilde{R} \left(1 - (\tilde{\nabla}\Phi)^2\right) + 8 (\tilde{\nabla}^2\Phi)^2 - 8 (\tilde{\nabla}^\alpha\tilde{\nabla}^\beta \Phi)^2 \Bigg]\nonumber \\
    && + \frac{\lambda\alpha_3}{2G} \int dx \sqrt{-\tilde{h}} \Bigg[ 4\tilde{K} \Big(1  - (\tilde{\nabla}\Phi)^2 + 2(\tilde{n}\cdot\tilde{\nabla}\Phi)^2 \Big) \nonumber \\
    && + 8\tilde{n}\cdot\tilde{\nabla}\Phi \left(\tilde{\nabla}_\alpha\tilde{\nabla}_\beta\Phi \tilde{n}^\alpha \tilde{n}^\beta - \tilde{\nabla}^2\Phi \right)\Bigg].
\end{eqnarray} 
\end{itemize}

The above equations correspond to the reduction of both, bulk and boundary actions. Here we notice that some of the higher derivative terms in the action will drop off when $2\tilde{\alpha}_1+\tilde{\alpha}_2$ vanishes, but nevertheless other higher derivative terms survive.

%where $\tilde{g}$ is two dimensional metric, $\tilde{R}$ is the two dimensional Ricci scalar. As in two-derivative case, dimensional reduction of an action with two fields, metric and the matter, produces an extra dilaton field $\Phi$ in the lower dimensions. As in two dimensions, the Riemann tensor has only one independent component and that can be thought of as the Ricci scalar. Then as speculated, all the higher derivative terms in the action in two dimensions, boils down to suitable higher powers of Ricci scalar and derivatives of $\Phi$.

\subsection{Classical Solutions of two dimensional field variables}
Solving equations of motions from the above two dimensional action, we find the classical solutions of the fields. A generic solution of the gauge field is given by,
\begin{align}
    & \tilde{F}_{\alpha\beta} = \frac{Q}{\Phi^2}\sqrt{-\tilde{g}}\ \varepsilon_{\alpha\beta},
\end{align}
Here, $Q$ is an integration constant which we identify with the charge of the higher dimensional black hole,
\begin{align}
    Q^2 = \Phi_0^2\left(1 + \frac{3\Phi_0^2}{L^2} \right).
\end{align}
We consider a constant solution for the dilaton. Using the gauge field solution, we get the value of the constant,
\begin{align}
    & \Phi = \tilde{\Phi}_0 \equiv \Phi_0 + \lambda (2\tilde{\alpha}_1 + \tilde{\alpha}_2) \frac{L^2 + 3\Phi_0^2}{L^2 + 6\Phi_0^2}\ \frac{6\Phi_0}{L^2}. \label{extdil} 
\end{align}
The scalar curvature takes a constant, negative value as
\begin{align}
    & \tilde{R} = -\frac{2}{\tilde{L}_2^2} = -2\left( \frac{1}{L_2^2} - \lambda (2\tilde{\alpha}_1 + \tilde{\alpha}_2) \frac{L^2 + 3\Phi_0^2}{L^2 + 6\Phi_0^2}\ \frac{12}{L^2\Phi_0^2} \right). \label{extmet}
\end{align}
These solutions with constant dilaton profile correctly capture the near horizon field configuration of the four-dimensional extremal black hole having $S^2$ and $AdS_2$ radii given by equations \eqref{extS2} and \eqref{extAdS2} respectively \cite{Kolekar:2018sba,Moitra:2019bub}. As in the four dimensional case, for $2\tilde{\alpha}_1 + \tilde{\alpha}_2=0,$ the extremal solution does not get any higher derivative corrections.

\subsection{A useful Weyl transformation} \label{Weyl}
As our goal is to get a JT-like theory in lower dimensions, we need to get rid of the kinetic term of the dilation $\Phi$. To do so, in this section, we perform a  Weyl re-scaling of the metric around the constant solution of the dilaton as,
\begin{align}
    \tilde{g}_{\alpha\beta} = \frac{\tilde{\Phi}_0}{\Phi} g_{\alpha\beta}.
\end{align}
As an artefact of this re-scaling the kinetic term of the dilaton disappears from the two derivative part of the action. This feature will be essential for the later part of the work and we shall come back to it in the next section. In terms of the re-scaled metric, the two-dimensional action \eqref{2D_act} takes the following form,
\begin{align}
    \bar{S} = \bar{S}^{(0)} + \bar{S}^{\Tilde{\alpha}} + \bar{S}^{GB}, \label{2DW}
\end{align}
where the respective parts of the action take the following forms,
%With the re-scaled metric, the parts of the action take the following forms:
\begin{itemize}
\item 
Two derivative part is
\begin{align}\label{2weyl}
    \bar{S}^{(0)} = & \frac{1}{4G} \int d^2x \sqrt{-g} \left( \Phi^2 R + \frac{2\tilde{\Phi}_0}{\Phi} -2\Lambda \Phi\tilde{\Phi}_0 - \frac{\Phi^3}{\tilde{\Phi}_0} F^2 \right) \nonumber \\
    & + \frac{1}{2G} \int dx \sqrt{-h} \Phi^2 \left(K + 2\frac{\Phi}{\tilde{\Phi}_0} n_\alpha F^{\alpha\beta} A_\beta\right).
\end{align}
\item Arbitrary higher derivative part has the following form
\begin{eqnarray}\label{4weylalpha}
    && \bar{S}^{\tilde{\alpha}}= \frac{\lambda}{4G} \int d^2x \sqrt{-g}\Bigg[\left(4\tilde{\alpha}_1+2\tilde{\alpha}_2\right)\frac{\Phi_0}{\Phi^3} + 4\tilde{\alpha}_1 R - \left(12\tilde{\alpha}_1+4\tilde{\alpha}_2\right)\left(\frac{\nabla^2\Phi}{\Phi} + \frac{(\nabla\Phi)^2}{\Phi^2}\right) \nonumber \\
    && + \frac{\Phi}{\Phi_0} \Bigg\{ \left(\tilde{\alpha}_1+\frac{\tilde{\alpha}_2}{2}\right) \Phi^2 R^2 - \left(6\tilde{\alpha}_1 + \tilde{\alpha}_2\right)\ (\Phi \nabla^2\Phi 
    + (\nabla\Phi)^2) R  + 4\tilde{\alpha}_2 \nabla^\alpha\nabla^\beta\Phi \Big(\nabla_\alpha\nabla_\beta\Phi\nonumber \\
    &&  + \frac{2\nabla_\alpha\Phi\nabla_\beta\Phi}{\Phi} \Big) +  \left(9\tilde{\alpha}_1+\frac{\tilde{\alpha}_2}{2}\right) (\nabla^2\Phi)^2 + \left(9\tilde{\alpha}_1+\frac{9\tilde{\alpha}_2}{2}\right)\frac{(\nabla\Phi)^4}{\Phi^2} \nonumber \\
    &&+ \left(18\tilde{\alpha}_1+\tilde{\alpha}_2\right) \frac{\nabla^2\Phi (\nabla\Phi)^2}{\Phi}  \Bigg\}\Bigg]  + \frac{\lambda}{2G} \int dx \sqrt{-h}\Bigg[\left(-\frac{24\tilde{\alpha}_1}{L^2}-\frac{6\tilde{\alpha}_2}{L^2} + \tilde{\alpha}_2 \frac{\Phi^2}{\Phi_0^2}F^2\right)\Phi^2 K \nonumber \\
    && + \left(-\frac{36\tilde{\alpha}_1}{L^2}-\frac{9\tilde{\alpha}_2}{L^2} + 4\tilde{\alpha}_2 \frac{\Phi^2}{\Phi_0^2} F^{\alpha\gamma}\tensor{F}{^\beta_\gamma} n_\alpha n_\beta - \frac{5\tilde{\alpha}_2}{2} \frac{\Phi^2}{\Phi_0^2}F^2\right)\Phi n\cdot\nabla\Phi\Bigg],
\end{eqnarray}
\item Whereas Gauss-Bonnet part is
\begin{eqnarray}\label{4weylGB}
    && \bar{S}^{GB} = \frac{\lambda\alpha_3}{4G} \int d^2x \sqrt{-g}\Bigg[ 4R + \frac{\Phi}{\Phi_0} \Bigg\{ - 4R(\nabla\Phi)^2  + \frac{4\nabla^2\Phi(\nabla\Phi)^2}{\Phi} + 8 (\nabla^2\Phi)^2 \nonumber \\
    &&  - 8\nabla^\alpha\nabla^\beta\Phi \Big(\nabla_\alpha\nabla_\beta\Phi+ \frac{2\nabla_\alpha\Phi\nabla_\beta\Phi}{\Phi} \Big)\Bigg\} \Bigg]  + \frac{\lambda\alpha_3}{2G} \int dx \sqrt{-h} \Bigg[4K + \frac{\Phi}{\Phi_0} \Bigg\{ - 4K(\nabla\Phi)^2 \nonumber \\
    &&+ 8K (n\cdot\nabla\Phi)^2  + n\cdot\nabla\Phi \left( 8\nabla_\alpha\nabla_\beta\Phi n^\alpha n^\beta - 8\nabla^2\Phi + \frac{4(n\cdot\nabla\Phi)^2}{\Phi} - \frac{2(\nabla\Phi)^2}{\Phi} \right) \Bigg\} \Bigg].\nonumber\\
\end{eqnarray}
\end{itemize}
The re-scaled action also admits a constant dilaton solution when the spacetime has constant negative curvature as given by \eqref{extdil} and \eqref{extmet}.

\subsection{Integrating out the abelian gauge field} 
%Thus once we use the above gauge field solution in the equations of motion of the metric and the dilaton, coming from the action \eqref{2DW}, we find that these equations can be found from an effective action with only metric and dilaton field, where the substitution of the gauge field solutions acts like potential term \cite{Kolekar:2018sba}. This could be easily checked at the two-derivative level,
\iffalse
\begin{eqnarray}
   S_{Max}+S_{Maxb}|_{on-shell} &=& \frac{1}{4G}\int d^2x \sqrt{-g}\frac{\Phi^3}{\tilde{\Phi}_0}F^2\nonumber\\
   &=&- \frac{1}{4G}\int d^2x \sqrt{-g}\frac{Q^2\tilde{\Phi}_0}{\Phi^3}
\end{eqnarray}
\fi

Since the gauge field in two dimensions does not have any dynamics, it can be integrated out\footnote{It is shown in appendix \ref{2dgauge} that gauge fields in two dimensions can be integrated out \cite{Witten:1991we} and the effective action depends on the choice of a measure, which in this case depends on the dilaton. Since the field strength is constant at the boundary, the higher derivative boundary terms do not contribute in the gauge field integration.} to obtain an effective action involving the metric and dilaton as given by:
\begin{align}
    S = S^{(0)} + S^{\tilde{\alpha}} + S^{GB}, \label{2D_eff}
\end{align}
where the respective parts of the action in Euclidean signature are,
\begin{itemize}
\item Two derivative part is
\begin{equation}
  S^{(0)}=  - \frac{1}{4G} \int d^2x \sqrt{g} \left( \Phi^2 R + \frac{2\tilde{\Phi}_0}{\Phi} -2\Lambda \Phi\tilde{\Phi}_0 - \frac{2Q^2\tilde{\Phi}_0}{\Phi^3} \right) - \frac{1}{2G} \int dx \sqrt{h} \Phi^2 K  
\end{equation}
\item Arbitrary part is given by
\begin{eqnarray}
    && S^{\tilde{\alpha}} = - \frac{\lambda}{4G} \int d^2x \sqrt{g}\Bigg[\left(4\tilde{\alpha}_1+2\tilde{\alpha}_2\right)\frac{\Phi_0}{\Phi^3} + 4\tilde{\alpha}_1 R - \left(24\tilde{\alpha}_1+8\tilde{\alpha}_2\right) \frac{(\nabla\Phi)^2}{\Phi^2} \nonumber \\
    && + \frac{\Phi}{\Phi_0} \Bigg\{ \left(\tilde{\alpha}_1+\frac{\tilde{\alpha}_2}{2}\right) \Phi^2 R^2 - \left(6\tilde{\alpha}_1 + \tilde{\alpha}_2\right)\ (\Phi \nabla^2\Phi 
    + (\nabla\Phi)^2) R  + 4\tilde{\alpha}_2 \nabla^\alpha\nabla^\beta\Phi \Big(\nabla_\alpha\nabla_\beta\Phi \nonumber \\
    &&+ \frac{2\nabla_\alpha\Phi\nabla_\beta\Phi}{\Phi} \Big)  +  \left(9\tilde{\alpha}_1+\frac{\tilde{\alpha}_2}{2}\right) (\nabla^2\Phi)^2 + \left(9\tilde{\alpha}_1+\frac{9\tilde{\alpha}_2}{2}\right)\frac{(\nabla\Phi)^4}{\Phi^2}\nonumber \\
    &&  + \left(18\tilde{\alpha}_1+\tilde{\alpha}_2\right) \frac{\nabla^2\Phi (\nabla\Phi)^2}{\Phi}  \Bigg\}\Bigg] - \frac{\lambda}{2G} \int dx \sqrt{h}\Bigg[ (-6\tilde{\alpha}_1 - 2\tilde{\alpha}_2)\frac{n\cdot\nabla\Phi}{\Phi} - \Big(\frac{24\tilde{\alpha}_1}{L^2}\nonumber\\
    &&+\frac{6\tilde{\alpha}_2}{L^2} + \frac{2\tilde{\alpha}_2 Q^2}{\Phi^4}\Big)\Phi^2 K  - \Big(\frac{36\tilde{\alpha}_1}{L^2}+\frac{9\tilde{\alpha}_2}{L^2} -  \frac{\tilde{\alpha}_2 Q^2}{\Phi^4}\Big)\Phi n\cdot\nabla\Phi\Bigg]
\end{eqnarray}
\item Gauss Bonnet part is
\begin{eqnarray}
    && S^{GB}= - \frac{\lambda\alpha_3}{4G} \int d^2x \sqrt{g}\Bigg[ 4R + \frac{\Phi}{\Phi_0} \Bigg\{ - 4R(\nabla\Phi)^2  + \frac{4\nabla^2\Phi(\nabla\Phi)^2}{\Phi} + 8 (\nabla^2\Phi)^2  \nonumber\\&&- 8\nabla^\alpha\nabla^\beta\Phi \Big(\nabla_\alpha\nabla_\beta\Phi + \frac{2\nabla_\alpha\Phi\nabla_\beta\Phi}{\Phi} \Big)\Bigg\} \Bigg]  - \frac{\lambda\alpha_3}{2G} \int dx \sqrt{h} \Bigg[4K + \frac{\Phi}{\Phi_0} \Bigg\{ - 4K(\nabla\Phi)^2\nonumber \\
    && + 8K (n\cdot\nabla\Phi)^2  + n\cdot\nabla\Phi \left( 8\nabla_\alpha\nabla_\beta\Phi n^\alpha n^\beta - 8\nabla^2\Phi + \frac{4(n\cdot\nabla\Phi)^2}{\Phi} - \frac{2(\nabla\Phi)^2}{\Phi} \right) \Bigg\} \Bigg]\nonumber\\
\end{eqnarray}
\end{itemize}

Thus we get a two dimensional effective theory for metric and scalar that describes the physics of four dimensional system around extremality.

\subsection{Fluctuation around extremality}
The constant dilaton solution \eqref{extdil} and \eqref{extmet} of the two-dimensional theory captures the near-horizon configuration of a spherically symmetric extremal black hole solution of the four dimensional theory \cite{Kolekar:2018sba,Moitra:2019bub}. Since we are interested in a near-extremal solution, we fluctuate the fields around their classical values \cite{Kolekar:2018sba}. 
In two dimensions, the metric has three independent components, two of which can be removed by choosing a particular gauge so that the dynamics of the two-dimensional metric is now captured by a scalar $\omega$. In this gauge, the metric takes the following form 
\begin{align}
    ds^2=\text{e}^{2\omega}(dt^2+dz^2).
\end{align}
Fluctuations around the classical solutions %\eqref{adsmet} and \eqref{dilsol}
are given by,
\begin{align}
    & \Phi=\tilde{\Phi}_0(1+\epsilon\phi), \\
    & \omega=\omega_0+\epsilon\Omega.
\end{align}
Here $\epsilon$ is a small parameter controlling the order of fluctuations and the background extremal solutions are, 
\begin{align}
    & \Phi = \tilde{\Phi}_0 , \\
    & \text{e}^{2\omega_0}=\frac{\tilde{L}_2^2}{z^2} .
\end{align}
We expand the full action \eqref{2D_eff} order by order in $\epsilon$, where the dynamical fields are $\Omega$ and $\phi$. The metric and related quantities in the following actions are evaluated for the background. 
\vspace{0.2cm}
The action at the order of $\epsilon^0$ is given by,
\begin{align}
    S_0  = & - \frac{1}{4G} \int d^2x \sqrt{g} \Bigg[ (\tilde{\Phi}_0^2 + \lambda\ 4\tilde{\alpha}_1 + \lambda\ 4\alpha_3) R \nonumber \\
    &  + \lambda (2\tilde{\alpha}_1 + \tilde{\alpha}_2) \left\{ \frac{24}{L^2}\left( 1 + \frac{3\Phi_0^2}{L^2} \right) + \frac{2}{\Phi_0^2} + \frac{1}{2}R^2\Phi_0^2\right\}  \Bigg] \nonumber \\
    & - \frac{1}{2G} \int dx \sqrt{h} \left( \tilde{\Phi}_0^2  - \lambda\ 2\tilde{\alpha}_2 + \lambda\ 4\alpha_3 - \lambda (2\tilde{\alpha}_1 + \tilde{\alpha}_2) \frac{12\Phi_0^2}{L^2} \right) K.
\end{align}
$S_0$ is a constant which corresponds to the extremal entropy \eqref{extent}.

Whereas the action at linear order of $\epsilon$,
\begin{align}
    S_1 = & - \frac{1}{4G} \int d^2x \sqrt{g} \Bigg[ 2\phi \tilde{\Phi}_0^2 \left( R + \frac{2}{\tilde{L}_2^2} \right) + \lambda (2\tilde{\alpha}_1 + \tilde{\alpha}_2)\left(\frac{1}{2}R^2\Phi_0^2- \frac{2\Phi_0^2}{L_2^4} \right)\ 3\phi \nonumber \\
    &  - \lambda (6\tilde{\alpha}_1 + \tilde{\alpha}_2) \Phi_0^2 R \nabla^2\phi \nonumber - 2\nabla^2\Omega  \left\{ \tilde{\Phi}_0^2 + \lambda4\tilde{\alpha}_1 + \lambda 4\alpha_3 + \lambda(2\tilde{\alpha}_1 + \tilde{\alpha}_2)\Phi_0^2R \right\} \nonumber \\
    &  + \lambda (2\tilde{\alpha}_1 + \tilde{\alpha}_2) \left\{\frac{4}{\Phi_0^2}-\Phi_0^2R^2+\frac{48}{L^4}(L^2+3\Phi_0^2) \right\}\Omega  \Bigg] \nonumber \\
    & - \frac{1}{2G} \int dx \sqrt{h} \Bigg[ 2\phi K \left( \tilde{\Phi}_0^2 + \lambda\ 2\tilde{\alpha}_2 - \lambda\frac{24\tilde{\alpha}_1\Phi_0^2}{L^2} \right) - \lambda (6\tilde{\alpha}_1 + \tilde{\alpha}_2) \frac{\Phi_0^2}{L_2^2} n\cdot\nabla\phi \nonumber \\
    &  + \left\{ \tilde{\Phi}_0^2 - 2\lambda\tilde{\alpha}_2 + \lambda 4\alpha_3 - \lambda(2\tilde{\alpha}_1 + \tilde{\alpha}_2)\frac{12\Phi_0^2}{L^2} \right\} n\cdot\nabla\Omega  \Bigg].
\end{align}
For the asymptotically $AdS_2$ background with length scale $\tilde{L}_2$, the  $\nabla^2\phi$ and $\nabla^2\Omega$ terms are total derivatives and they exactly cancel with the boundary terms $n\cdot\nabla\phi$ and $n\cdot\nabla\Omega$ respectively. Rest of the terms in the bulk expression i.e. the $\phi$ and $\Omega$ potential terms are identically zero. The resulting linear action is independent of the metric fluctuation $\Omega$ and it reduces to a boundary term only, as shown below
%
\iffalse
\begin{eqnarray}
   && S_1 =  - \frac{1}{4G} \int d^2x \sqrt{g} \Bigg[ 2\phi \tilde{\Phi}_0^2 \left( R + \frac{2}{\tilde{L}_2^2} \right) + \lambda (2\tilde{\alpha}_1 + \tilde{\alpha}_2)\left(\frac{1}{2}R^2\Phi_0^2- \frac{2\Phi_0^2}{L_2^4} \right)\ 3\phi\Bigg] \nonumber \\
    && - \frac{1}{2G} \int dx \sqrt{h} \Bigg[ 2\phi K \left( \tilde{\Phi}_0^2 + \lambda\ 2\tilde{\alpha}_2 - \lambda\frac{24\tilde{\alpha}_1\Phi_0^2}{L^2} \right)\Bigg].
\end{eqnarray}
{\color{Red}{The bulk term vanishes for the background value of the metric. Thus the linear action reduces to the boundary term only, as shown below }}
\fi
%
\begin{align}
    S_1 = - \frac{\tilde{\Phi}_0^2}{G} \int_{\phi=\phi_b} dx \sqrt{h}\left( 1 + \lambda\frac{2\tilde{\alpha}_2}{\Phi_0^2} - \lambda\frac{24\tilde{\alpha}_1}{L^2} \right) \phi K. \label{bdyaction}
\end{align}
Equations of motion of the fields $\Omega$ and $\phi$ can be obtained from the $\epsilon^2$ action. Solving these equations we get a time-independent solution for $\phi$ as the following
\begin{align}\label{solphi}
    \phi = \frac{A_1}{z} + A_2 z^2,
\end{align}
where $A_1$ ans $A_2$ are constants. The first term in \eqref{solphi} is linearly increasing towards the boundary, whereas the second term dies off very fast, hence irrelevant for our purposes. 

\subsection{Boundary action from a modified JT action}
In this section, we find the higher derivative  modification of the JT theory that reproduces the same boundary dynamics as action \eqref{bdyaction}. Since we are considering small fluctuations around extremality, we expect this action to be linear in the scalar. A subclass of the action \eqref{4derdil}, which is linear in $\phi$, is given by, 
\begin{align}
    S = -\frac{1}{16\pi G_2}\int d^2x \sqrt{g}\Bigg[ \phi \left(R+\frac{2}{L_2^2}\right) + \lambda V\phi + \lambda Z\phi R^2  \Bigg]. \label{modJT}
\end{align}
Here, $V$ has dimension of inverse length square whereas $Z$ has dimension of length square which we shall choose accordingly. For $\lambda=0$, the action is that of the  two derivative JT theory and it describes the near horizon dynamics of a near-extremal black hole in Einstein-Hilbert-Maxwell theory. The $\phi$ e.o.m. sets the scalar curvature to a constant. We want the metric to be $AdS_2$ with radius $\tilde{L}_2$ given by \eqref{extAdS2}\footnote{It is the near-horizon $AdS_2$ radius of the four-dimensional near-extremal black hole.}. Demanding the e.o.m. of $\phi$ to be $R = -\frac{2}{\tilde{L}_2^2}$, we fix the constant $V$,
\begin{align}
    V = - (2\tilde{\alpha}_1 + \tilde{\alpha}_2) \frac{L^2 + 3\Phi_0^2}{L^2 + 6\Phi_0^2}\ \frac{24}{L^2\Phi_0^2} - \frac{4Z}{L_2^4}.
\end{align}
The action thus becomes,
\begin{align}
    S = -\frac{1}{16\pi G_2}\int d^2x \sqrt{g}\Bigg[ \phi \left(R+\frac{2}{\tilde{L}_2^2}\right) + \lambda Z\phi\left( R^2 - \frac{4}{L_2^4}\right) \Bigg]. 
\end{align}
For a well-defined variational principle with Dirichlet boundary conditions on the fields we add an appropriate boundary term following the similar method of \cite{Cremonini:2009ih},
\begin{align}
    S_{GH} = -\frac{1}{8\pi G_2}\int dx\sqrt{h} \left(1-\lambda \frac{4Z}{L_2^2}\right)\phi K.\label{modJT_bdy}
\end{align}
Comparing with the linear action \eqref{bdyaction}, we find,
\begin{align}
    & G_2 = \frac{G}{8\pi\tilde{\Phi}_0^2} \label{G2}, \\
    & Z = \frac{12\tilde{\alpha}_1\Phi_0^2 - \tilde{\alpha}_2 L^2}{2(L^2 + 6\Phi_0^2)}.
\end{align}
The full action is given by\footnote{In addition, a counter-term should be added such that the effective action remains finite as $\varepsilon\rightarrow 0$. Since the boundary curvature is zero, the counter-term differs from that of a two-derivative theory by a constant scaling \cite{Cremonini:2009ih}.},
\begin{align}
    S = & -\frac{1}{16\pi G_2}\int d^2x \sqrt{g}\Bigg[ \phi \left(R+\frac{2}{\tilde{L}_2^2}\right) + \lambda \frac{12\tilde{\alpha}_1\Phi_0^2 - \tilde{\alpha}_2 L^2}{2(L^2 + 6\Phi_0^2)}\left( R^2 - \frac{4}{L_2^4}\right)\phi \Bigg] \nonumber \\ 
    &-\frac{1}{8\pi G_2}\int dx\sqrt{h} \left( 1 + \lambda\frac{2\tilde{\alpha}_2}{\Phi_0^2} - \lambda\frac{24\tilde{\alpha}_1}{L^2} \right)\phi K. \label{modiJT} 
\end{align}
The dilaton has a linearly varying solution when the metric is $AdS_2$.

Similar to the JT gravity action, the dilaton acts as a Lagrange multiplier in the higher derivative corrected action \eqref{modiJT}. The Euclidean path integral computation for this higher derivative corrected theory can be performed similarly as in subsection \ref{EPI}. In the corresponding Euclidean path integral, the dilaton can be integrated out along an imaginary line with Dirichlet boundary condition. This results in a delta function constraint on the metric configurations such that the fields having non-zero contribution to the integral must satisfy,
\begin{align}
    \left(R+\frac{2}{\tilde{L}_2^2}\right) + \lambda \frac{12\tilde{\alpha}_1\Phi_0^2 - \tilde{\alpha}_2 L^2}{2(L^2 + 6\Phi_0^2)}\left( R^2 - \frac{4}{L_2^4}\right) = 0
\end{align}
Since the higher derivative terms are of perturbative nature, we need to find the scalar $R$ that solves the above equation perturbatively in $\lambda$. For $\lambda = 0$, this equation corresponds to $R + 2/L_2^2 = 0$. Using this the full equation can be solved iteratively to $\lambda$-order and we get,
\begin{align}
    R+\frac{2}{\tilde{L}_2^2}=0.
\end{align}
These geometries are locally $AdS_2$ with length $\tilde{L}_2$, having wiggly boundary curves $\mathcal{C}: (t(u),z(u)) = (t(u),\varepsilon t'(u))$ and the boundary conditions are,
\begin{align} 
    & g|_\mathcal{C} = \frac{\tilde{L}_2^2}{\varepsilon^2} \label{metricbc}, \\
    & \phi|_\mathcal{C} = \frac{\phi_b}{\varepsilon} \label{dilatonbc}. 
\end{align}

The boundary value $\phi_b$ is identified with the location of the boundary in the action \eqref{bdyaction}. Using these boundary conditions, the effective action becomes a Schwarzian of the boundary modes. The path integral reduces to sum over these modes $t(u)$ with $SL(2,R)$ identification, 
\begin{align}
    Z(\beta) =  \int \frac{d\mu(t)}{SL(2,R)}\text{exp}\left[\tilde{\phi}_b\int^{\beta}_0 du\ \text{Sch}[t(u),u]\right], \label{modSch}
\end{align}\\
where, $\beta$ is the periodicity of the boundary time coordinate and the coupling constant $\tilde{\phi}_b$ for the boundary action is given by,
\begin{align}
    \tilde{\phi}_b = \bar{\phi}_b\left( 1 + \lambda\frac{2\tilde{\alpha}_2}{\Phi_0^2} - \lambda\frac{24\tilde{\alpha}_1}{L^2} \right). \label{mod_coup}
\end{align}
Here $\bar{\phi}_b = \frac{\phi_b}{8\pi G_2}$ is the coupling \eqref{sch_coup} of the unmodified boundary theory.

\subsection{Thermodynamic quantities for small temperature}
Since the near-horizon low energy dynamics is governed by an effective boundary action, the free energy and other thermodynamic quantities of the near-extremal black hole can be obtained from the Euclidean partition function corresponding to the boundary theory. It is evident from \eqref{modSch} that the effect of the higher derivative corrections is included as a multiplicative constant in the boundary action. Hence the integration can be performed exactly like \eqref{SchPI}. Following the method reviewed in section \ref{NE_stat}, we obtain the partition function,
\begin{align}
    Z_{JT}(T) \simeq \left(\tilde{\phi}_b T\right)^{3/2}\text{exp}\left(2\pi^2\tilde{\phi}_b T\right).
\end{align}
The parameter $T = \frac{1}{\beta}$ is identified with the temperature\footnote{The temperature can be written in terms of unmodified temperature $T_0$ as discussed in \eqref{temp}.} of the near-extremal black hole in the higher derivative corrected Einstein-Maxwell theory. Comparing with the four-dimensional black hole solution \eqref{RN}, we can identify the $S^2$ radius $\Phi(r) = r$. The near horizon $AdS_2$ factor of the black hole solution can be put into the form \eqref{adsmet}, following the transformation,
\begin{align}
    z = \frac{\tilde{L}_2^2}{r - \tilde{\Phi}_0}.
\end{align}
From the definition of the dilaton fluctuation,
\begin{align}
    \phi = \frac{r - \tilde{\Phi}_0}{ \tilde{\Phi}_0} = \frac{\tilde{L}_2^2}{\tilde{\Phi}_0 z}.
\end{align}
From this relation, we identify the boundary value of the dilaton,
\begin{align}
    & \phi_b = \frac{\tilde{L}_2^2}{\tilde{\Phi}_0} \label{dilbdy}.
\end{align}
Using the expression of Newton's constant \eqref{G2} and the boundary value of dilaton \eqref{dilbdy}, we get the coupling \eqref{mod_coup},
\begin{align}
    \tilde{\phi}_b = \frac{\tilde{L}_2^2\tilde{\Phi}_0}{G}\left( 1 + \lambda\frac{2\tilde{\alpha}_2}{\Phi_0^2} - \lambda\frac{24\tilde{\alpha}_1}{L^2} \right).
\end{align}

The thermodynamic quantities above extremality can be computed from the expressions of section \ref{NE_stat} in terms of the modified coupling $\tilde{\phi}_b$ and temperature $T$, 
\begin{align}
    & \delta F  = -2\pi^2 \tilde{\phi}_b T^2 - \frac{3}{2}T\log\left({\tilde{\phi}_b T}\right), \\
    & \delta S = \frac{3}{2} + 4\pi^2 \tilde{\phi}_b T + \frac{3}{2}\log\left({\tilde{\phi}_b T}\right), \label{NE_ent} \\
    & \delta M  = 2\pi^2 \tilde{\phi}_b T^2 + \frac{3}{2}T. \label{NE_energy} 
    \end{align}
In absence of higher derivative corrections i.e. for $\lambda = 0$, we have
\begin{align*}
    & \tilde{\phi}_b = \bar{\phi}_b = \frac{L_2^2\Phi_0}{G}, \\
    & T = T_0,
\end{align*}
where $L_2$ and $\Phi_0$ are the unmodified $AdS_2$ and $S^2$ radii of the extremal black hole and $T_0$ is the unmodified temperature of a near-extremal black hole with the same charge.

The first term in \eqref{NE_energy} comes from the saddle point contribution to the partition function, whereas the second term arises due to exact computation of the path integral. As stated in the introduction, a naive semiclassical analysis would imply that the 
%mass above extremality for small enough temperatures $T<M_{gap}$\footnote{Under saddle point approximation, the mass above extremality is of the form $\delta M \sim \frac{T^2}{M_{gap}}$, with $M_{gap}\equiv 1/\bar{\phi}_b$.}, is smaller than the average energy of Hawking radiation which is linear in temperature i.e. 
the black hole cannot radiate even though it has a small finite temperature and  there exists a mass gap $M_{gap}$ between an extremal black hole and a near-extremal black hole (with the smallest temperature) having the same charge. It was argued in \cite{Iliesiu:2020qvm} that in this regime, the semiclassical description breaks down since the fluctuations around the classical solution have considerable contribution to the partition function and hence these quantum corrections should be taken into account. The quantum fluctuations result in logarithmic corrections to free energy and entropy and it implies that there is no mass gap in the spectrum because the black hole energy is always greater than the average energy of radiation. The constant $M_{gap}$ is rather related to the scale of $SL(2,R)$ symmetry breaking. In our computation we find that the entropy of a near-extremal black hole in higher derivative modified theory also has a similar logarithmic correction. This is one of the prime results of this paper.

\section {JT equivalence of 4D near-extremal near-horizon dynamics with more generic four derivative interactions} \label{gen4derth}
So far we have only considered four derivative metric interaction terms in the four dimensional theory. In general, there are other possible four derivative corrections involving gauge field e.g. $RF^2, R_{\mu\nu\rho\sigma}F^{\mu\nu}F^{\rho\sigma}, (F^2)^2, \tensor{F}{^\mu_\nu}\tensor{F}{^\nu_\rho}\tensor{F}{^\rho_\sigma}\tensor{F}{^\sigma_\mu}$ and covariant derivatives of gauge field strength e.g. $\nabla_\rho F_{\mu\nu} \nabla^\rho F^{\mu\nu}, F_{\mu\nu}\nabla^\nu\nabla_\rho F^{\mu\rho}$, with arbitrary coupling constants. In this section we briefly discuss the effects of such corrections to our results. Let us first understand how the near horizon extremal geometry get modified due to presence of such generic four derivative interactions. We note that,
\begin{itemize}
    \item The extremal background will get modified due to corrections having no covariant derivatives of the gauge field strength. 
    \item Owing to the $SO(2,1)\times SO(3)$ near horizon symmetry, the gauge field strength and scalars take constant values near the horizon of an extremal black hole. Due to the form of the near horizon field configuration, covariant derivatives of the fields vanish in this background \cite{Sen:2007qy}. Hence, the extremal solution will get no contribution from corrections involving covariant derivatives of the gauge field strength.
\end{itemize}

For electrically charged solutions, after dimensional reduction, we will have a theory of gravity coupled to a scalar and gauge field like before. The constant scalar solution in this dimensionally reduced theory will capture the near-horizon configuration of the (un)modified extremal solution. Unlike our previous considerations, the gauge field solution will be modified. However in two dimensions, the gauge field strength has to be proportional to the Levi-Civita symbol. Hence, we can obtain an effective theory involving metric and dilaton where the gauge field terms will act as potential terms for the dilaton, depending on the charge. In this effective theory, we can consider small fluctuations around the extremal background and these fluctuations describe the near-extremal dynamics. The linearized action should reduce to a boundary action like \eqref{bdyaction} with a pre-factor depending on the coefficients of the higher derivative interactions.

The remaining analysis of thermodynamics is identical to that of section \ref{sec4}.
The near-extremal dynamics can also be derived from a modified JT action \eqref{modJT} with suitable choices of the constants $G_2, \phi_b, V,$ and $Z$ such that effectively it becomes the equivalent of the boundary action like \eqref{bdyaction}. The constants will clearly depend on the coefficients of the four derivative corrections which can be evaluated by explicit computations. The corresponding entropy above extremality, equivalent to  \eqref{NE_ent} can also be computed as before.

Thus we find that the near-horizon dynamics of a near-extremal black hole at low energy in a four dimensional theory having arbitrary four derivative corrections is captured by a JT-like action \eqref{modJT} that contains a $R^2$ correction only. The scalar equation of motion in such a theory gives a constant negative curvature solution with a length scale given by the near horizon $AdS_2$ factor of the four dimensional extremal black hole. Effectively, this theory can be described by a boundary action like \eqref{modJT_bdy}. With appropriate boundary conditions, this boundary term becomes a Schwarzian derivative of time reparametrization modes. We find that the $R^2$ corrected JT theory shares all the aspects of the JT model since the coupling of the $R^2$ interaction contributes to a modification to the coupling constant of the boundary theory only, which just modifies the symmetry breaking scale by a small amount. We expect this equivalence between a near-extremal black hole in a higher dimensional theory with arbitrary four derivative corrections and a $R^2$ corrected JT action \eqref{modJT} to hold even when started from dimension $D>4$ with more generic four derivative terms.

\section{Conclusions and Open Problems} \label{concl}
In this paper we have studied the  most generic four derivative corrected four dimensional near-extremal black hole. We find that all the thermodynamic characteristics of this class of black holes are captured in a particular four derivative ($R^2$) corrected JT gravity theory. The dynamics of this higher derivative corrected JT theory is given by a boundary Schwarzian action. In particular, we show that the Schwarzian captures the near-extremal entropy above extremality of the corresponding black hole. The spectrum does not have a mass gap at low enough temperature and it is consistent with the modified semi-classical analysis. 

Let us quickly recall the importance of the above study. In the context of understanding the statistical source of black hole entropy,
it was shown long back in \cite{Dabholkar:2004yr, Sen:2004dp} that if we take into account a class of higher derivative
corrections to the effective action that describes the black hole, the entropy of the black hole
solution reproduces the statistical entropy. This was particularly important for the small black hole solutions that get a zero entropy at the level two derivative gravity theory.
The higher derivative corrections to the effective theory are capable of stretching the horizon of a small black
hole in arbitrary dimensions and thus produces a non zero entropy \cite{Sen:2005kj} for them. Thus higher derivative corrections play a fundamental role in understanding black hole entropy. In the present work, the analysis is performed for the near-extremal black holes. We have shown that the higher derivative corrected thermodynamics of the near-extremal black hole physics can be suitably incorporated in a higher derivative modified JT theory. The equivalence of this thermodynamic entropy to that of a statistical entropy still remains an open question.   

A byproduct of our computations is the modifications of the $M/Q$ ratio at fixed charge $Q$. The near-extremal mass to charge ratio can be found using \eqref{NE_energy}, which is a new result of this paper. For very low temperatures, the second term in \eqref{NE_energy} dominates, which is obtained from an exact path integral computation. Depending on the higher derivative modifications coming from the coupling of boundary theory and the temperature, the $M/Q$ ratio\footnote{The result at extremality is given in \eqref{MR}, which is in agreement with the results of \cite{Cremonini:2019wdk}. For a particular class of 4D near-extremal solution, the ratio in classical regime is given in \eqref{MRNE}.} changes. This change can have implications on possible ranges of the higher-derivative couplings of the theory, based on the Weak Gravity Conjecture \cite{Cremonini:2019wdk, Arkani-Hamed:2021ajd}. This study is out of the scope of this present paper, but nevertheless is an interesting one to look for.  

Let us finish the paper with possible open directions.
JT serves as toy model in various studies. It has been seen that JT and JT-like two derivative dilaton gravity model is dual to random matrix model. There is study on the Freudenthal duality of near extremal black hole entropy in supergravity in the framework of JT gravity \cite{TMAC}. It would be interesting to find the implications of the higher derivative correction terms on these dualities. Gravitational deformation of JT can be thought as $T\bar{T}$ deformation of JT coupled with matter \cite{Ishii:2019uwk}. It would also be interesting to generalize this study in the higher derivative modified JT theories. 

Another interesting direction to pursue is to compare our present results of near-extremal black hole entropy with other ways of computing the thermodynamics of non-extremal black holes. In particular such a result was given in \cite{Sen:2012dw}. It would be interesting to find the correspondence between these two apparently different constructions at two derivative theory and to extend it further to higher the derivative case. We would report on this comparison in future.

\section*{Acknowledgement}
We would like to thank Shamik Banerjee, Suvankar Dutta and Upamanyu Moitra for helpful discussions. The work is partially supported by SERB ECR grant, Govt. of India. The work of TM is supported by the Simons Foundation Grant Award ID 509116 and by the South African Research Chairs initiative of the Department of Science and Technology and the National Research Foundation. Finally we acknowledge the support of people of India towards fundamental research. 

\appendix
\section{Most generic dilaton gravity action in two dimensions}
\label{gendilact}
\subsection{Two derivative action}
The most generic form of dilaton gravity theory in two dimensions having at most two derivatives is given by, 
\begin{equation}
   S = -\frac{1}{16\pi G_2} \int d^2x \sqrt{g}\left[ U_1(\Phi) R + U_2(\Phi)\nabla_\mu\Phi\nabla^\mu\Phi + U_3(\Phi) \right]. \label{gendil}
\end{equation}
Two out of the three functions $U_1,U_2,U_3$ can be eliminated \cite{Witten:2020ert}. Assuming $U'_1(\Phi)\neq 0$, the scalar can be redefined as, 
\begin{align}
    \phi = U_1(\Phi).
\end{align}
In terms of the redefined field, the action \eqref{gendil} takes the form
\begin{align}
    S = -\frac{1}{16\pi G_2} \int d^2x \sqrt{g}\left[ \phi R + \tilde{U}_2(\phi)\nabla_\mu\phi\nabla^\mu\phi + \tilde{U}_3(\phi) \right].
\end{align}
The function $\tilde{U}_2(\phi)$ can be eliminated by redefining the metric using a Weyl transformation of the following form, 
\begin{align}
    & g_{\mu\nu}\rightarrow \text{exp}\left(2\sigma(\phi)\right)g_{\mu\nu}.
\end{align}
such that $2\sigma'(\phi)=-\tilde{U}_2(\phi)$. Finally, the action reduces to
\begin{equation}
    S = -\frac{1}{16\pi G_2}\left[ \int d^2x \sqrt{g}\left( \phi R + V(\phi)\right) + 2\int dx \sqrt{h} \phi K  \right],
\end{equation}   
where the boundary term is added to satisfy the variational principle. 

\subsection{Four-derivative action}
The most generic four-derivative corrected dilaton gravity action is given by \cite{Elizalde:1993ga},
\begin{eqnarray}\label{gen4action}
   && S = -\frac{1}{16\pi G_2}\int d^2x \sqrt{g}\Bigg[  \phi R + V(\phi) + \lambda Z_1(\phi) (\nabla\phi)^4 + \lambda Z_2(\phi) \nabla_\mu\nabla_\nu\phi\nabla^\mu\phi\nabla^\nu\phi \nonumber \\
    && + \lambda Z_3(\phi)\nabla_\mu\nabla_\nu\phi\nabla^\mu\nabla^\nu\phi + \lambda Z_4(\phi) R(\nabla\phi)^2 +  \lambda Z_5(\phi) R\nabla^2\phi +  \lambda Z_6(\phi) R^2  \Bigg],
\end{eqnarray}
where $Z_1, Z_2, Z_3, Z_4, Z_5$ and $Z_6$ are arbitrary functions of the dilaton $\phi$. $\lambda$ is a dimensionless parameter controlling the strength of four derivative terms. Since the number of independent components of the Riemann tensor in two spacetime dimensions is one and that can be thought of as Ricci scalar $R$, only $R^2$, $R (\nabla \phi)^2$, $R (\nabla^2 \phi)$ and other four derivative combinations of the derivatives of $\phi$ appear in the action.
Some of the four-derivative terms can be absorbed in the two-derivative part of the action using the following field redefinition
\begin{eqnarray}\label{fldredf}
    g_{\mu\nu} \rightarrow & &\left[ 1 + \lambda k_1(\phi)\nabla^2\phi + \lambda k_2(\phi)(\nabla\phi)^2 + \lambda k_3(\phi) R \right]g_{\mu\nu} \nonumber \\
  && + \lambda k_4(\phi)\nabla_\mu\nabla_\nu\phi + \lambda k_5(\phi)\nabla_\mu\phi\nabla_\nu\phi.
\end{eqnarray}
Under this field redefinition, the two-derivative part $\sqrt{g}(\phi R + V(\phi))$ transforms as,
\begin{align*}
    & \sqrt{g}(\phi R + V(\phi)) + \lambda\sqrt{g}\left(\frac{V}{2}-\nabla^2\phi\right)\left[(2k_1+k_4)\nabla^2\phi + (2k_2+k_5) (\nabla\phi)^2 + 2k_3 R\right] \nonumber \\
   & +  \lambda\sqrt{g}\nabla^\mu\nabla^\nu\phi\left[ (k_1\nabla^2\phi + k_2(\nabla\phi)^2 +  k_3 R) g_{\mu\nu} + k_4\nabla_\mu\nabla_\nu\phi + k_5\nabla_\mu\phi\nabla_\nu\phi\right].
\end{align*}
We want to absorb higher-derivative terms using the field redefinition, keeping the form of the two-derivative part invariant i.e. we do not want any two-derivative correction to $\sqrt{g}(\phi R + V(\phi))$. From this demand, we get
\begin{align*}
    & 2k_1(\phi)+k_4(\phi) = 0, \\
    & 2k_2(\phi)+k_5(\phi) = 0.
\end{align*}
We further redefine the dilaton $\phi \rightarrow \phi + \lambda k_3(\phi)V(\phi)$. The transformation of $\sqrt{g}(\phi R + V(\phi))$ term 
\begin{align*} 
    \sqrt{g}& \Bigg[\phi R + V(\phi) +  \lambda k_1(\phi)(\nabla^2\phi)^2 +  \lambda k_2(\phi)\nabla^2\phi(\nabla\phi)^2 - 2\lambda k_1(\phi)\nabla_\mu\nabla_\nu\phi\nabla^\mu\nabla^\nu\phi \nonumber \\
    & - 2\lambda k_2(\phi)\nabla_\mu\nabla_\nu\phi\nabla^\mu\phi\nabla^\nu\phi -\lambda k_3(\phi)  R\nabla^2\phi \Bigg].
\end{align*}

The $k_1(\nabla^2\phi)^2$ and $k_2\nabla^2\phi(\nabla\phi)^2$ terms can be rewritten as linear combinations of the six higher derivative terms of \eqref{gen4action} up to total derivatives
\begin{align*}
    & k_2\nabla^2\phi(\nabla\phi)^2 = -k'_2(\nabla\phi)^4 - 2k_2\nabla_\mu\nabla_\nu\phi\nabla^\mu\phi\nabla^\nu\phi, \\
    & k_1(\nabla^2\phi)^2 = k''_1(\nabla\phi)^4 + 3k'_1\nabla_\mu\nabla_\nu\phi\nabla^\mu\phi\nabla^\nu\phi + k_1\nabla_\mu\nabla_\nu\phi\nabla^\mu\nabla^\nu\phi + \frac{k_1}{2}R(\nabla\phi)^2.
\end{align*}
Therefore, the transformed two-derivative part of \eqref{gen4action} takes the following form
\begin{align}
    \sqrt{g}\Bigg[& \phi R + V + \lambda\Bigg((k''_1-k'_2)(\nabla\phi)^4 + (3k'_1-4k_2)\nabla_\mu\nabla_\nu\phi\nabla^\mu\phi\nabla^\nu\phi \nonumber \\
    & - k_1\nabla_\mu\nabla_\nu\phi\nabla^\mu\nabla^\nu\phi  + \frac{k_1}{2}R(\nabla\phi)^2 - k_3R\nabla^2\phi\Bigg) \Bigg]. \label{redf2d}
\end{align}
Plugging in the above expression \eqref{redf2d} in the action \eqref{gen4action}, three out of six four-derivative terms can be removed by choosing the arbitrary functions $k_1,k_2,$ and $k_3$. One such choice is,
\begin{align}
    k_1 = Z_3, \quad k_2 = \frac{1}{4}(Z_2+Z'_3), \quad k_3 = Z_5.
\end{align}
The four-derivative dilaton gravity action depending on four arbitrary functions can be written as,
\begin{align}\label{4derdilapp}
    S = -\frac{1}{16\pi G_2}\int d^2x & \sqrt{g}\Bigg[ \phi R + V(\phi) + \lambda W_1(\phi) (\nabla\phi)^4 + \lambda W_2(\phi) R(\nabla\phi)^2 + \lambda W_3(\phi) R^2  \Bigg].
\end{align}

\section{Variational principle in the presence of higher derivative corrections}

We consider the four-dimensional Einstein-Maxwell action
\begin{align}
    S^{(0)}_{bulk} = \frac{1}{16\pi G}\int d^4x \sqrt{-g}\left(R - 2\Lambda - F_{\mu\nu}F^{\mu\nu}\right).
\end{align}
The action contains two derivatives of the metric. Taking small variations of the metric leads to $\delta g_{\mu\nu}$ terms and its derivatives at the boundary. If we consider Dirichlet boundary conditions on the metric i.e. if we hold the metric fixed at the boundary, the variations and their tangential derivatives vanish at the boundary but normal derivatives at the boundary survive. To make the variation of the action zero, we need to add the Gibbons-Hawking boundary term at the boundary, given by
\begin{align}
    S_{GH} = \frac{1}{8\pi G}\int d^3x \sqrt{-h} K.
\end{align}
For negative cosmological constant $\Lambda= - \frac{3}{L^2}$, the vacuum solution is $AdS$ spacetime which has a timelike boundary at spatial infinity. Here $h_{\mu\nu}= g_{\mu\nu} - n_\mu n_\nu$ is the induced metric and $n_\mu$ is the normal to the boundary. $K=\nabla_\mu n^\mu$ is the extrinsic curvature. 

We add a generic Gauss-Bonnet correction to the bulk action given by,
\begin{align}
    \frac{1}{16\pi G}\int d^4x \sqrt{-g}\lambda\left(\alpha_1 R^2 + \alpha_2 R_{\mu\nu}R^{\mu\nu}+\alpha_3 R_{\mu\nu\rho\sigma}R^{\mu\nu\rho\sigma}\right).
\end{align}
The correction term contains four-derivative terms of the metric. In general, it is difficult to obtain an exact boundary term for this action. But the Gauss-Bonnet combination $R_{GB}^2 = R^2 - 4 R_{\mu\nu}R^{\mu\nu}+ R_{\mu\nu\rho\sigma}R^{\mu\nu\rho\sigma}$ keeps the equations of motion\footnote{In four dimensions, Gauss-Bonnet combination is a topological invariant and hence it does not contribute to the equations of motion.} of second order and the boundary term can be generalized for this particular combination. Hence for simplicity we rewrite the correction as,
\begin{align}
    \frac{1}{16\pi G}\int d^4x \sqrt{-g}\lambda\left(\tilde{\alpha}_1 R^2 + \tilde{\alpha}_2 R_{\mu\nu}R^{\mu\nu}+\alpha_3 R_{GB}^2\right).
\end{align}
Here $\tilde{\alpha}_1 = \alpha_1-\alpha_3$ and $\tilde{\alpha}_2 = \alpha_2+4\alpha_3$. For $\tilde{\alpha}_1=\tilde{\alpha}_2=0$, the Gibbons-Hawking boundary term can be generalized \cite{Myers:1987yn},
\begin{align}
    S_{GH}^{GB} = \frac{\lambda\alpha_3}{4\pi G}\int d^3x \sqrt{-h} (J - 2G^{(3)}_{\mu\nu}K^{\mu\nu}),
\end{align}
Here $J$ is the trace of $J_{\mu\nu} = \frac{1}{3} (2{K}K_{\mu \rho}K^\rho_\nu + K_{\rho\sigma}K^{\rho\sigma}K_{\mu\nu} -2K_{\mu \rho}K^{\rho\sigma}K_{\sigma\nu} -K^2 K_{\mu\nu})$ and $G^{(3)}_{\mu\nu}$ is the Einstein tensor corresponding to the induced metric. 

However for $\tilde{\alpha}_1\neq 0$ and $\tilde{\alpha}_2\neq 0$, the equations of motion are higher order so that an exact boundary term cannot be found. Since we will treat the higher derivative terms as perturbative corrections to the two-derivative action, we can obtain a boundary term for which the total variation is $\mathcal{O}(\lambda^2)$. In this case, we use the leading equation of motion: 
\begin{align}
    R_{\mu\nu} = -\frac{3}{L^2} g_{\mu\nu} + 2F_{\mu\rho}\tensor{F}{_\nu^\rho} -\frac{1}{2}g_{\mu\nu}F^2.
\end{align} 
The boundary term is given by \cite{Cremonini:2009ih},
\begin{align}
  \frac{\lambda}{8\pi G}\int d^3x \sqrt{-h}\left[\tilde{\alpha}_2 (-KF^2+2F^{\mu\rho}\tensor{F}{^\nu_\rho}(Kn_\mu n_\nu + K_{\mu\nu})) -\frac{6}{L^2}\left(4\tilde{\alpha}_1+\tilde{\alpha}_2 \right)K\right].   
\end{align}
For the gauge field, we cannot choose Dirichlet boundary condition since the boundary term contains gauge field strength. Hence we choose a boundary condition $\delta (n_\mu F^{\mu\nu}) = 0$ which corresponds to fixing electric charge. We need an additional boundary term for the gauge field part of the bulk action. It is the Maxwell boundary term
\begin{align}
    \frac{1}{8\pi G} \int d^3x \sqrt{-h} F^{\mu\nu}n_\mu A_\nu.
\end{align}

\section{Computation of attractor values using entropy function} \label{entfn}
We consider a spherically symmetric, electrically charged, extremal black hole solution in a theory described by the action \eqref{4Dbulk}. The near-horizon field configuration consistent with the $SO(2,1)\times SO(3)$ isometry of the $AdS_2\times S^2$ geometry is given by \cite{Sen:2007qy},
\begin{eqnarray}
   ds^2 &=& v_1\left(-r^2 dt^2 + \frac{dr^2}{r^2}\right) + v_2 (d\theta^2 + \sin^2\theta d\varphi^2) \label{NHmetric}, \\
   F_{rt} &=& e.
\end{eqnarray}
Integrating the lagrangian density over the sphere for this field configuration
\begin{align}
   f(v_1,v_2,e) =& \frac{1}{2G}(v_1-v_2) + \frac{3v_1v_2}{2GL^2} + \frac{1}{2G}\frac{v_2}{v_1}e^2 \nonumber \\
   + & \frac{\lambda}{2G}(2\tilde{\alpha}_1 + \tilde{\alpha}_2)\left(\frac{v_1}{v_2}+\frac{v_2}{v_1}\right) - \frac{2\lambda}{G}(\tilde{\alpha}_1+\alpha_3).
\end{align}
From the definition of electric charge $Q=Gq$, we identify
\begin{align*}
    & \frac{\partial f}{\partial e} = q, \\
    & e = \frac{Qv_1}{v_2}.
\end{align*}
Taking Legendre transformation of $f$ with respect to the variable $e$ we get the entropy function $\mathcal{E}(v_1,v_2,q) = 2\pi (eq - f)$, where
\begin{align}\label{ent_func}
   \mathcal{E}(v_1,v_2,q) = & \frac{\pi}{G}\frac{v_1}{v_2}Q^2-\frac{\pi}{G}(v_1-v_2) -\frac{3\pi}{G}\frac{v_1v_2}{L^2} \nonumber \\
   - & \frac{\lambda\pi}{G}(2\tilde{\alpha}_1 + \tilde{\alpha}_2)\left(\frac{v_1}{v_2}+\frac{v_2}{v_1}\right) + \frac{4\lambda}{G}(\tilde{\alpha}_1+\alpha_3).
\end{align}
Extremizing the entropy function with respect to $v_1$ and $v_2$ we get the attractor values. Without the higher derivative corrections ($\lambda=0$), we get the $S^2$ and $AdS_2$ radii
\begin{align}
   & v_2|_{\lambda=0} \equiv \Phi_0^2 = \frac{L^2}{6}\left(\sqrt{1+\frac{12Q^2}{L^2}}-1\right), \\
   & v_1|_{\lambda=0} \equiv L_2^2 = \frac{L^2}{6}\left(1-\frac{1}{\sqrt{1+\frac{12Q^2}{L^2}}}\right).
\end{align}
Treating the higher derivative terms perturbatively, we obtain the modified $S^2$ and $AdS_2$ radii,
\begin{align}
    & v_2 = \tilde{\Phi}_0^2 = \Phi_0^2 + \lambda(2\tilde{\alpha}_1 + \tilde{\alpha}_2)\frac{12Q^2/L^2}{\sqrt{1+\frac{12Q^2}{L^2}}}, \\
    & v_1 = \tilde{L}_2^2 = L_2^2 + \lambda(2\tilde{\alpha}_1 + \tilde{\alpha}_2)\frac{12Q^2/L^2}{(1+\frac{12Q^2}{L^2})^{3/2}}.
\end{align}
The entropy of the extremal black hole is given by the value of the entropy function at this extrema 
\begin{align}
    S_{ext} = \mathcal{E}(\tilde{L}_2^2,\tilde{\Phi}_0^2,Q) = \frac{\pi\Phi_0^2}{G}-\lambda(\tilde{\alpha}_2-2\alpha_3)\frac{2\pi}{G} - \lambda(2\tilde{\alpha}_1 + \tilde{\alpha}_2)\frac{36\pi\Phi_0^4}{GL^2(L^2+6\Phi_0^2)}. \label{extent}
\end{align} 

\section{Integrating out gauge fields in 2D} \label{2dgauge}
In this appendix, we shall integrate out the gauge field from the theory described in section \ref{Weyl} following \cite{Witten:1991we}. The gauge field dependent bulk part of the action \eqref{2DW} is as follows,
\begin{align}
    S_g = -\frac{1}{4G}\int d^2x \sqrt{-g} \frac{\Phi^3}{\tilde{\Phi}_0}F^2 = \frac{1}{2G}\int F \wedge *F\label{gaugeact}.
\end{align}

In 2D, $F$ must be proportional to the volume form $\Omega$ since it is a two-form and the Hodge dual $*F$ is a scalar: 
\begin{align}
    F = f\Omega, \qquad \Omega = \frac{1}{2}\frac{\tilde{\Phi}_0}{\Phi^3}\sqrt{g} \epsilon_{\mu\nu}dx^\mu \wedge dx^\nu.
\end{align}
In Euclidean signature, \eqref{gaugeact} can be written in terms of the scalar $f$ as,
\begin{align}
    S_g = -\frac{1}{2G}\int \Omega f^2 \label{YM_measure}.
\end{align}
We consider an action involving the gauge field and an auxiliary scalar $X$ such that it is linear in gauge field,
\begin{align}
    S = -\frac{G}{2}\int \Omega X^2 - i\int X F \label{YM_aux}.
\end{align}
It can be readily seen from the path integral by performing a Gaussian integral for the auxiliary scalar that the action \eqref{YM_aux} is equivalent to the action \eqref{YM_measure}. It can also be obtained by putting the $X$ equation of motion back into \eqref{YM_aux} which is,
\begin{align}
    f = iGX \label{aux_def}.
\end{align}
The gauge field can be easily integrated out using \eqref{YM_aux} since it is linear in field strength. Using $X F=d(X A) - dX A$, \eqref{YM_aux} takes the following form,
\begin{align}
    S = -\frac{G}{2}\int \Omega X^2 + i\int dX A - i\int_{\partial} X A.
\end{align}
The boundary term cancels with the Maxwell boundary term. The gauge field dependent part\footnote{The gauge field strength terms at the boundary \eqref{4weylalpha} at order $\lambda$ do not affect the integration since gauge field strength is constant at the boundary.} of the path integral can be integrated out easily,
\begin{align}
    \int \mathcal{D}A \text{e}^{-i\int dX A} = \delta(dX).
\end{align}
The $X$ integral now gets contribution from $dX = 0$ only, which fixes the value of $X$ to a constant $X_b$. Using \eqref{aux_def}, we relate this constant to the charge $Q$ from asymptotic boundary conditions,
\begin{align}
    X = X_b = -i\frac{Q}{G}.
\end{align}
The corresponding effective action depends on the measure $\Omega$ and acts as a potential term for the field $\Phi$ in the full action,
\begin{align}
    S_{eff} = \frac{Q^2}{2G}\int d^2x \sqrt{g}\frac{\tilde{\Phi}_0}{\Phi^3}.
\end{align}

\section{Spherically symmetric charged black hole solution in four dimensions}
A spherically symmetric electrically charged black hole in the four dimensional Einstein-Maxwell theory with generic Gauss-Bonnet correction, described by the action \eqref{4Dbulk1}, has the following form
\begin{align}
    & ds^2 = -f_1(r)dt^2 + \frac{dr^2}{f_2(r)} + r^2 d\Omega^2, \label{RN} \\
    & f_1(r) = 1-\frac{2M}{r}+\frac{Q^2}{r^2} + \frac{r^2}{L^2} + \lambda\tilde{\alpha}_1\frac{24Q^2}{L^2r^2}+ \lambda\tilde{\alpha}_2 \left(\frac{2Q^2}{L^2r^2} - \frac{2Q^2}{r^4} +\frac{2MQ^2}{r^5} - \frac{2Q^4}{5r^6} \right), \\
    & f_2(r) = 1-\frac{2M}{r}+\frac{Q^2}{r^2} + \frac{r^2}{L^2} + \lambda\tilde{\alpha}_1\frac{24Q^2}{L^2r^2}+ \lambda\tilde{\alpha}_2 \left(\frac{6MQ^2}{r^5} - \frac{4Q^2}{r^4} - \frac{12Q^4}{5r^6} \right), \\
    & F_{rt} = \frac{Q}{r^2} + \lambda \tilde{\alpha}_2 \frac{Q^3}{r^6}.
\end{align}
The parameters $M, Q, L$ have dimensions of length and the higher derivative coefficients $\tilde{\alpha}_1, \tilde{\alpha}_2$ have dimensions of length square.
For an extremal black hole, the horizon is located at
\begin{align}
    r_h = \tilde{\Phi}_0 = \Phi_0 + \lambda(2\tilde{\alpha}_1 + \tilde{\alpha}_2)\frac{L^2+3\Phi_0^2}{L^2+6\Phi_0^2}\frac{6\Phi_0}{L^2}.
\end{align}
$\Phi_0$ is the extremal horizon without higher derivative corrections and it is related to the charge by,
\begin{align}
    & Q^2 = \Phi_0^2\left(1+\frac{3\Phi_0^2}{L^2}\right).
\end{align}
At extremality, the mass is given by,
\begin{align}
    & M_{ext} = \Phi_0 + \frac{2\Phi_0^3}{L^2} - \lambda(L^2+3\Phi_0^2)\frac{\tilde{\alpha}_2 L^2 - 12(5\tilde{\alpha}_1 + \tilde{\alpha}_2)\Phi_0^2}{5L^4\Phi_0}.
\end{align}
%The mass to charge ratio at extremality is for fixed charge is given by,
For fixed charge black hole, the mass to charge ratio at extremality is given by
\begin{align}\label{MR}
    \frac{M_{ext}}{Q} = \left(\frac{M_{ext}}{Q}\right)_{\lambda=0}\left[1+\lambda\frac{L^2+3\Phi_0^2}{5(L^2+2\Phi_0^2)}\left((5\tilde{\alpha}_1 + \tilde{\alpha}_2)\frac{12}{L^2}-\frac{\tilde{\alpha}_2}{\Phi_0^2}\right)\right].
\end{align}
The corresponding unmodified ratio is,
\begin{align}
    \left(\frac{M_{ext}}{Q}\right)_{\lambda=0} = \frac{L^2+2\Phi_0^2}{L\sqrt{L^2+3\Phi_0^2}}.
\end{align}
The above ratio is greater than one, as an artefact of the AdS geometry. In the presence of the higher derivative modification, the ratio gets modified as in  \eqref{MR}.
Keeping the charge fixed, if we slightly increase the mass above extremality, the extremal horizon splits into two close but distinct horizons. The horizons are given by
\begin{align}
    r_{\pm} = \tilde{\Phi}_0 \pm \Delta,
\end{align}
such that $f_1(r_{\pm})=f_1(r_{\pm})=0$. The mass above extremality is proportional to $\Delta^2$. The temperature of the near-extremal black hole is defined as,
\begin{align}
    T = \frac{1}{4\pi} \sqrt{f'_1(r_+)f'_2(r_+)}.
\end{align}
The temperature gets modified due to the higher derivative corrections. The corrected temperature is given by,
\begin{align}
    T = T_0 - \lambda T_0 \frac{L^2+3\Phi_0^2}{2L^2\Phi_0^2(L^2+6\Phi_0^2)^2}(\tilde{\alpha}_2 L^4 + 6(6\tilde{\alpha}_1 + 5\tilde{\alpha}_2)L^2\Phi_0^2 + 72(\tilde{\alpha}_1 + \tilde{\alpha}_2)\Phi_0^4), \label{temp}
\end{align}
where $T_0$ is the temperature of the near-extremal black hole in the absence of higher derivative corrections. The shift in horizon can be expressed in terms of $T_0$,
\begin{align}
    \Delta = T_0 \frac{2\pi L^2 \Phi_0^2}{L^2+6\Phi_0^2} & \Bigg(1  -  \lambda\frac{3(L^2+3\Phi_0^2)}{2L^2\Phi_0^2(L^2+6\Phi_0^2)^2}(\tilde{\alpha}_2 L^4 +  2(-2\tilde{\alpha}_1 + 5\tilde{\alpha}_2)L^2\Phi_0^2 + 24(\tilde{\alpha}_1 + 2\tilde{\alpha}_2)\Phi_0^4) \Bigg).
\end{align}
If instead, we fix the temperature to $T_0$ even under the presence of higher derivative corrections, the near extremal mass is given by,
\begin{align}
    M_{NE} = & M_{ext} + \frac{2\pi^2 L^2 \Phi_0^3}{L^2+6\Phi_0^2} T_0^2 \nonumber \\
    & + \lambda T_0^2\frac{2\pi^2 \Phi_0(L^2+3\Phi_0^2)}{(L^2+6\Phi_0^2)^3}(\tilde{\alpha}_2 L^4 +  6(6\tilde{\alpha}_1 + 5\tilde{\alpha}_2)L^2\Phi_0^2 + 72(\tilde{\alpha}_1 + \tilde{\alpha}_2)\Phi_0^4)
\end{align}
The shift in horizon is given as,
\begin{align}
    \Delta = T_0 \frac{2\pi L^2 \Phi_0^2}{L^2+6\Phi_0^2} \left(1 - \lambda \frac{(L^2+3\Phi_0^2)}{L^2\Phi_0^2(L^2+6\Phi_0^2)^2} (\tilde{\alpha}_2 L^4 - 24\tilde{\alpha}_1 L^2\Phi_0^2 + 36\tilde{\alpha}_2 \Phi_0^4)\right)
\end{align}
Mass to charge ratio of the near-extremal black hole solution,
\begin{align}\label{MRNE}
    \frac{M_{NE}}{Q} = \left(\frac{M_{NE}}{Q}\right)_{\lambda=0} \Bigg[ 1 + & \lambda\frac{L^2+3\Phi_0^2}{5(L^2+2\Phi_0^2)}\left((5\tilde{\alpha}_1 + \tilde{\alpha}_2)\frac{12}{L^2}-\frac{\tilde{\alpha}_2}{\Phi_0^2}\right) \nonumber \\
    & + \lambda T_0^2 \frac{4\pi^2L^2(L^2+3\Phi_0^2)}{5(L^2+2\Phi_0^2)^2(L^2+6\Phi_0^2)^3}\big(3\tilde{\alpha}_2 L^6 + 20L^4(3\tilde{\alpha}_1+4\tilde{\alpha}_2)\Phi_0^2 \nonumber \\
    & +276\tilde{\alpha}_2L^2\Phi_0^4+144(-5\tilde{\alpha}_1+\tilde{\alpha}_2)\Phi_0^6\big)\Bigg] .
\end{align}

\section{Kaluza-Klein reduction on \texorpdfstring{$S^2$}{text}} \label{KK}
In general, an infinite tower of modes appear in lower dimensions upon the dimensional reduction of a theory over a compact manifold. For the reduction of Einstein-Maxwell theory over a sphere, the lower dimensional theory can be consistently truncated to the massless sector consisting of finite number of modes \cite{pope}. 

We are interested in studying the near-horizon dynamics of a spherically symmetric near-extremal black hole in four-dimensional Einstein-Hilbert-Maxwell theory under the presence of four derivative interactions from two-dimensional perspective. The higher dimensional fields (i.e. metric and $U(1)$ gauge field) can be decomposed in terms of spherical harmonics $Y_{lm}(y)$, where ${y^i}$ denote the coordinates on $S^2$ and the labels $l,m$ correspond to $(2l+1)$ dimensional representation of $SO(3)$ with $-l\leq m \leq +l$ \cite{Michelson:1999kn}.  

The massless fields generated through the dimensional reduction are two dimensional metric, dilaton, corresponding to the radius of the sphere, $U(1)$ and $SO(3)$ gauge fields. The 2D metric, dilaton, and $U(1)$ gauge field are constant on the sphere and these fields correspond to the $l=0$ sector of the decomposition. At $l=1$ sector, only massless fields are the $SO(3)$ gauge fields which transform covariantly under the diffeomorphisms on $S^2$. For $l \geq 2$, all the fields are massive. Therefore, the dimensional reduction ansatz for consistent Kaluza-Klein reduction in massless sector has the following form,
\begin{align}
    & ds^2 = g_{\mu\nu}dx^\mu dx^\nu + \Phi^2(x^\alpha) g_{ij} (dy^i + K^i_m \mathcal{A}^m_\mu dx^\mu)(dy^j + K^j_n \mathcal{A}^n_\nu dx^\nu), \\
    &\hat{A}_{\mu} = \tilde{A}_\mu, \quad \hat{A}_i = 0. 
\end{align}
Here, ${x^\mu}$ are the coordinates on the 2D manifold with metric $g_{\mu\nu}$ and the metric on $S^2$ is $g_{ij}$, which has an $SO(3)$ isometry group generated by the Killing vectors $K^i_m$. $\mathcal{A}^m$ are the $SO(3)$ gauge fields and $\tilde{A}_\mu$ is the $U(1)$ gauge field. 

Following \cite{Witten:1991we, Iliesiu:2020qvm}, the $2D$ gauge fields can be integrated out to give an effective action for the metric and dilaton. This effective theory can also be simply obtained by plugging in the equations of motion for the gauge fields in the action. In this theory, the effects of the gauge fields are realized as a potential term for the dilaton depending on the $U(1)$ charge $Q$ and $SO(3)$ charges $l(l+1)$. In the presence of higher derivative corrections, after integrating out the gauge fields, we should get various interaction terms involving the metric and the dilaton which will also depend on the charges $Q$ and $SO(3)$ charges $l(l+1)$. 

In the effective theory, the constant dilaton solution with $l=0$ describes the near-horizon structure of a spherically symmetric extremal black hole. Since we are interested in the thermodynamics of a spherically symmetric near-extremal black hole with the same charges as the extremal solution, we can safely work with the $l=0$ sector modes only which leads to choosing the dimensional reduction ansatz \eqref{S^2_red}. Nevertheless, if we uplift a solution with non-zero values of $l$ from two to four dimensions, it will correspond to a rotating black hole. Therefore the $SO(3)$ modes are important in the study of the thermodynamics of Kerr-Newman black holes at low temperature.

\end{document}